\documentclass[acmsmall,screen,nonacm]{acmart}
\AtBeginDocument{%
  }
\usepackage{enumitem}
\usepackage{subfigure}
\usepackage{amsmath,amsfonts}
\usepackage{algorithmic}
\usepackage{graphicx}
\usepackage{textcomp}
\usepackage{xcolor}
\usepackage{booktabs}
\usepackage{tikz,pgfplots}
\usepackage{standalone}
\usepackage[most]{tcolorbox}
\usepackage{csquotes}
\usepackage{url}
\usepackage{multirow}
\usetikzlibrary{arrows.meta}
\usepackage{fancyhdr}
\usepackage{threeparttable}
\usepackage{subcaption}
\usepackage{makecell}
\usepackage{hyperref}
\usepackage{newtxmath}
\usepackage{pgf-pie}
\usepackage{multirow}
\usepackage{fontawesome5}
\usepackage{pifont} % Provides check marks and crosses
\usepackage{xspace}
\usepackage{tabularx}
\usepackage{adjustbox}
\usepackage{xr-hyper}
\usepackage{hyperref}

\usepackage[table]{xcolor}

\usepackage{tikz}
\usepackage{xcolor}
\usetikzlibrary{positioning,shapes,arrows.meta}

\usetikzlibrary{arrows.meta,calc,positioning}

\definecolor{beaublue}{rgb}{0.74, 0.83, 0.9}
\definecolor{applegreen}{rgb}{0.55, 0.71, 0.0}
\definecolor{dollarbill}{rgb}{0.52, 0.73, 0.4}
\definecolor{ao(english)}{rgb}{0.0, 0.5, 0.0}
\definecolor{amaranth}{rgb}{0.9, 0.17, 0.31}
\definecolor{cerublue}{rgb}{0.16, 0.32, 0.75}
\definecolor{frenchblue}{rgb}{0.0, 0.45, 0.73}
\definecolor{iceberg}{rgb}{0.44, 0.65, 0.82}
\definecolor{cadmiumorange}{rgb}{0.93, 0.53, 0.18}
\definecolor{myred}{rgb}{0.898,0.725,0.710}
\definecolor{myorange}{rgb}{0.988,0.922,0.867}
\definecolor{myyellow}{rgb}{0.984,0.980,0.855}
\definecolor{mygreen}{rgb}{0.922,0.945,0.875}

\newcommand{\leveldist}{1cm}        % Distance between levels
\newcommand{\siblingdist}{1.2cm}   % Distance between siblings
\newcommand{\yshiftone}{2cm}       % Vertical shift for first level
\newcommand{\yshifttwo}{1.5cm}     % Vertical shift for second level
\newcommand{\yshiftthree}{0.5cm}   % Vertical shift for third level

\newcommand{\bend}{0.5cm}

\tikzset{
  arrowstyle/.style={
    thick,
    -{Stealth[length=6pt,width=6pt]},
    rounded corners=0pt,
    shorten >=1pt
  }
}

\newcommand{\cmark}{\textcolor{applegreen}{\ding{51}}} % Green check mark
\newcommand{\xmark}{\textcolor{amaranth}{\ding{55}}} % Red cross mark

\makeatletter
\tcbset{
    myhbox/.style 2 args={%
        enhanced, 
        breakable,
        colback=white,
        colframe=cerublue!70,
        attach boxed title to top left={yshift*=-\tcboxedtitleheight}, 
        title={#2},
        boxed title size=title,
        boxed title style={%
            sharp corners, 
            rounded corners=northwest, 
            colback=tcbcolframe, 
            boxrule=0pt,
        },
        underlay boxed title={%
            \path[fill=tcbcolframe] (title.south west)--(title.south east) 
                to[out=0, in=180] ([xshift=5mm]title.east)--
                (title.center-|frame.east)
                [rounded corners=\kvtcb@arc] |- 
                (frame.north) -| cycle; 
        },
        #1
    }
}   
\makeatother

\newtcolorbox{myhbox}[2][]{%
    myhbox={#1}{#2}
}

\newcounter{findingcount}
\newcounter{suggestioncount}
\newcommand{\keybox}[1]{
\begin{tcolorbox}[leftrule=1mm,rightrule=1mm,toprule=0mm,bottomrule=0mm,left=1pt,right=2pt,top=2pt,bottom=2pt, colback=cadmiumorange!30]
\em #1
\end{tcolorbox}
}

\lstdefinelanguage{JavaScript}{
  keywords={typeof, new, true, false, catch, function, return, null, switch, var, if, in, while, do, else,
    case, break, let, const, yield, await, async, class, constructor, export, import, extends},
  keywordstyle=\color{blue}\bfseries,
  ndkeywords={boolean, throw, implements, import, this},
  ndkeywordstyle=\color{magenta}\bfseries,
  identifierstyle=\color{black},
  sensitive=false,
  comment=[l]{//},
  morecomment=[s]{/*}{*/},
  commentstyle=\color{green!50!black}\ttfamily,
  stringstyle=\color{red}\ttfamily,
  morestring=[b]',
  morestring=[b]"
}

\newcommand{\cparagraph}[1]{\vspace{0.5mm}\noindent\textbf{#1}}

\setcopyright{acmlicensed}
\copyrightyear{2024}
\acmYear{2024}
\acmDOI{XXXXXXX.XXXXXXX}

\acmJournal{CSUR}
\acmISBN{978-1-4503-XXXX-X/18/06}

\begin{document}

\title{AI Agentic Programming: A Survey of Techniques, Challenges, and Opportunities}
\author{Huanting Wang}
\email{H.Wang7@leeds.ac.uk}
\orcid{0000-0002-9829-913X}
\affiliation{%
  \institution{University of Leeds}
  \city{Leeds}
  \country{UK}
}

\author{Jingzhi Gong}
\email{J.Gong@leeds.ac.uk}
\orcid{0000-0003-4551-0701}
\affiliation{%
  \institution{University of Leeds}
  \city{Leeds}
  \country{UK}
}

\author{Huawei Zhang}
\email{schz@leeds.ac.uk}
\orcid{0000-0003-1021-3184}
\affiliation{%
  \institution{University of Leeds}
  \city{Leeds}
  \country{UK}
}

\author{Jie Xu}
\email{J.Xu@leeds.ac.uk}
\orcid{0000-0002-4598-167X}
\affiliation{%
  \institution{University of Leeds}
  \city{Leeds}
  \country{UK}
}

\author{Zheng Wang}
\email{z.wang5@leeds.ac.uk}
\orcid{0000-0001-6157-0662}
\affiliation{%
  \institution{University of Leeds}
  \city{Leeds}
  \country{UK}
}

\renewcommand{\shortauthors}{Wang et al.}





\begin{abstract}
% The abstract should be at most 100 words long and consist of short, direct sentences. 
AI agentic programming is an emerging paradigm where large language model (LLM)-based coding agents autonomously plan, execute, and interact with tools such as compilers, debuggers, and version control systems. Unlike conventional code generation, these agents decompose goals, coordinate multi-step processes, and adapt based on feedback, reshaping software development practices. This survey provides a timely review of the field, introducing a taxonomy of agent behaviors and system architectures and examining relevant techniques for planning, context management, tool integration, execution monitoring, and benchmarking datasets. We highlight challenges of this fast-moving field and discuss opportunities for building reliable, transparent, and collaborative coding agents.

\end{abstract}

 \maketitle

\section{Introduction}
\label{sec:introduction}

The software development paradigm is changing rapidly with the rise of large language models~(LLMs)~\cite{hou2024largelanguagemodelssoftware}. These models enable artificial intelligence (AI) systems that not only translate natural language descriptions into code snippets~\cite{TOSEM25} but also understand task requirements, interact with development tools, and iteratively refine their outputs to produce complex software~\cite{FSE24_Alshahwan, qian2024chatdevcommunicativeagentssoftware}.  Recent studies suggest that developers now use LLMs routinely to assist in daily coding tasks~\cite{hou2024largelanguagemodelssoftware, DBLP:conf/fose-ws/FanGHLSYZ23, gong2025language}.
Unlike traditional code generation tools~\cite{TraditionalCodeGenration} that respond to a single prompt with a static code snippet, emerging AI coding agents are designed to operate within dynamic software environments, performing iterative, tool-augmented tasks to achieve complex goals.

This shift has given rise to a new programming paradigm, \textbf{AI agentic programming}, where LLM-based coding agents can autonomously plan, execute, and refine software development tasks~\cite{gao2023retrieval, novikov2025alphaevolve}. Unlike conventional code-completion tools~\cite{IntelliSense,LSP,FSE09_Bruch}, which primarily assist with local suggestions, these agents are capable of generating entire programs or modules from natural language specifications, diagnosing and fixing bugs using compiler or test feedback, writing and executing test cases, and refactoring code for readability or performance. Beyond code generation, they can also orchestrate external tools, such as compilers, debuggers, performance profilers, and version control systems, supporting an end-to-end development workflow.

This emerging paradigm has the potential to fundamentally change how software is built and maintained. For example, an AI agent can take a natural language description of a feature and work through a series of steps, such as writing code, generating tests, running those tests, analyzing and fixing issues, and preparing a pull request~\cite{qian2024chatdevcommunicativeagentssoftware}.  Some state-of-the-art coding agents, like Anthropic's Calude Opus 4~\cite{Calude4},   have demonstrated the ability to continue working for hours while maintaining task consistency, avoiding deadlocks, and recovering from failed actions~\cite{qian2024chatdevcommunicativeagentssoftware,novikov2025alphaevolve}. These systems can generate and test code, migrate software between frameworks, debug runtime failures, and integrate new features by decomposing complex goals into manageable subtasks~\cite{deng2024pentestgptllmempoweredautomaticpenetration,liu2023bolaa}. This represents a clear shift from static, one-shot AI-based code generation to \textit{interactive, iterative, and tool-augmented workflows}.

Although progress has been fast, AI agentic programming is still in its early stages. Existing systems vary in architecture, autonomy, tool integration, and reasoning capabilities. There is no standard taxonomy, benchmark suite, or evaluation methodology. At the same time, multiple key challenges remain. These include ensuring reliability and robustness in dynamic environments~\cite{hou2024largelanguagemodelssoftware}, mitigating errors and hallucinations in generated code~\cite{gong2025language}, extending support beyond dominant languages such as Python to diverse platforms and software ecosystems~\cite{DBLP:conf/pldi/polycoder22}, and embedding safety, trust, and accountability into autonomous behaviours~\cite{yang2024swe}.  

The success of AI coding agents also depends heavily on their ability to interact effectively with external tools. However, today's programming languages, compilers, and debuggers are fundamentally human-centric~\cite{DLS23_Marron, Aho_compilers}. They are not designed for automated, autonomous systems. These tools often abstract away internal states and decision-making processes to improve usability, ensure portability, and reduce cognitive load for human users~\cite{llvm, Maleki}. While this abstraction benefits human developers, it may not fit AI agents, which require fine-grained, structured access to internal states, transformation sequences, and validation logic in order to reason about the effects of their actions~\cite{bi2024iterative}.
Without such access, AI agents struggle to diagnose failures, understand the implications of their changes, or recover from errors in a principled way. For instance, when a code transformation leads to a build failure, the agent needs more than just an error message - it must trace the failure to specific intermediate steps and understand why certain code edits or actions caused the issue. Existing development environments do not provide hooks and feedback mechanisms to support this kind of iterative, tool-integrated reasoning. Similarly, agentic coding systems would benefit from toolchains that support \textit{iterative development}, \textit{state tracking}, and \textit{rich feedback propagation} - capabilities that most conventional tools do not expose. To operate effectively, AI agents may need access to internal compiler representations, transformation traces, symbolic information, and execution metadata. 

%This raises a fundamental question: \textit{Are our current programming languages and software development tools still adequate in the era of AI agentic programming?} Or is it time to rethink the design of programming languages, compilers, and debuggers to treat AI agents as first-class participants in the development process?

These challenges show that AI agentic programming is not just a new way of using existing tools. It is a shift that exposes important gaps in how today's systems software is designed.  As the field evolves rapidly, there is an urgent need to clarify its conceptual landscape, identify common patterns and system architectures, and assess the suitability of current development ecosystems.  This is the right moment to step back, take stock of recent progress, and lay out the key questions that researchers and developers need to tackle next.
Therefore, this survey aims to provide a comprehensive overview of the emerging field of AI agentic programming. Specifically, it covers:
\vspace{-2mm}
\begin{itemize}[leftmargin=*] 
\item A conceptual foundation and taxonomy of AI coding agents,
\item A review of core system architectures and underlying techniques,
\item A summary of the behavior dimensions, operating modes, evaluation strategies and benchmarking practices of AI coding agents,
\item A discussion of key challenges and current limitations, and
\item An exploration of future research directions, including opportunities to bridge perspectives across disciplines such as programming languages, software engineering, AI, and human-computer interaction.
\end{itemize}

We focus primarily on \textbf{LLM-driven agentic systems for software development}, though many insights extend to general AI agents in other domains like information retrieval~\cite{li2025matching}. Our goal is to chart the current landscape, clarify foundational concepts, and support the design of robust, efficient, and trustworthy AI agents for programming.
\section{Background}
\label{sec:background}
\subsection{AI Agentic Programming}

\begin{figure}[t!]
    \centering
    \includegraphics[width=0.6\linewidth]{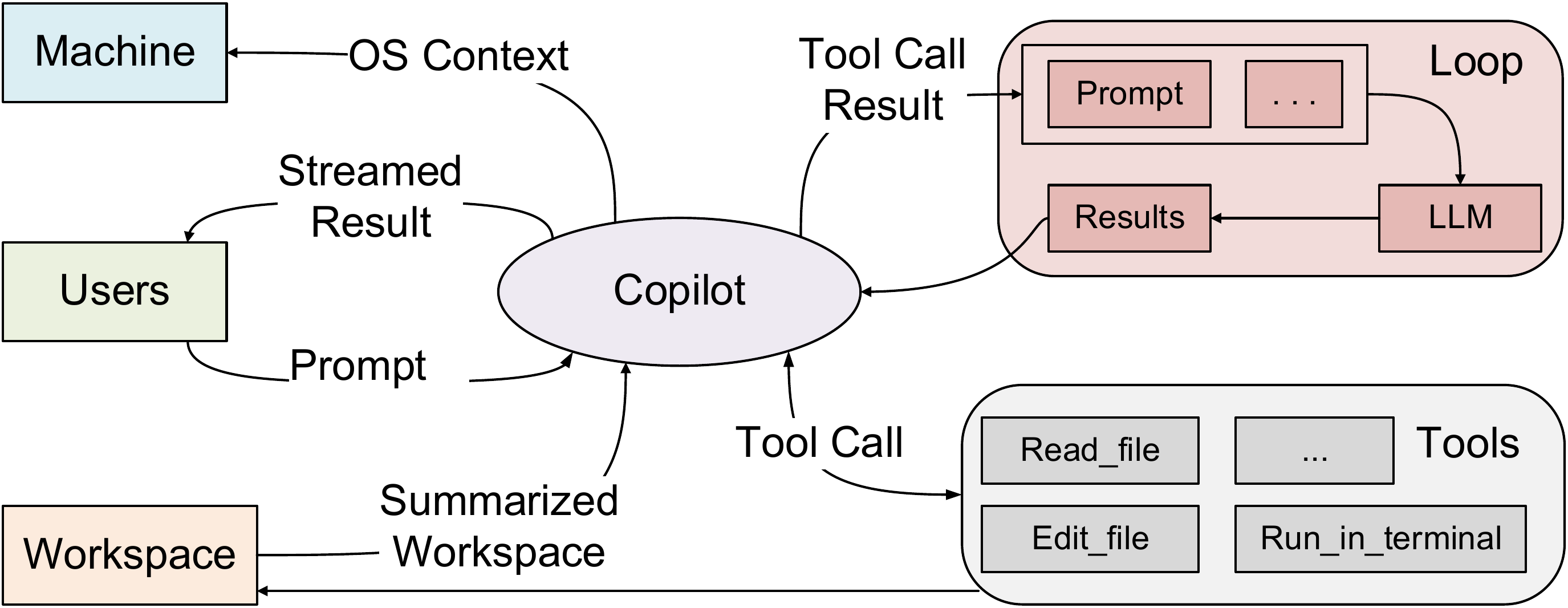}
    \caption{An example workflow of an AI coding agent.}
    \label{fig:cagent}
\end{figure}

AI agentic programming refers to a new programming paradigm in which LLM-based agents autonomously perform software development tasks. Unlike traditional code generation tools that produce outputs in a single step based on a static prompt \cite{DBLP:conf/pldi/polycoder22}, agentic systems operate in a goal-directed, multi-step manner. They reason about tasks, make decisions, use external tools (such as compilers, debuggers, and test runners), and iteratively refine their outputs based on feedback \cite{DBLP:conf/nips/ShinnCGNY23, liu2024large, novikov2025alphaevolve}. These agents can plan sequences of actions, adapt their strategies over time, and coordinate complex development workflows with limited or no human intervention.

At its core, agentic programming combines the capabilities of natural language processing, external tool integration, and task planning. Figure~\ref{fig:cagent} illustrates the architecture of a GitHub Copilot-style agentic programming system \cite{github_copilot}. At its core, the agent embeds an LLM within an execution loop, enabling interaction with the development environment. The LLM receives natural language prompts from the user and gathers additional context from the operating system and the workspace (e.g., file summaries or environment state). This information is passed into the reasoning loop, where the LLM decomposes the task into subgoals, generates code or decisions, and determines whether to invoke external tools like reading/editing files or executing terminal commands. Tool outputs are returned to the loop and used as feedback for further refinement. This iterative process continues until the agent completes the task or reaches a stopping condition. Final results are streamed back to the user.

A typical AI agentic programming system is characterized by several key properties. First, it emphasizes \textit{autonomy}, where LLM-based agents can make decisions and take actions without continuous human supervision. Second, it is inherently \textit{interactive}, as agents engage with external tools and environments during execution. Third, it supports \textit{iterative refinement}, allowing agents to improve outputs based on intermediate feedback. Importantly, it is \textit{goal-oriented}, with agents pursuing high-level objectives (e.g., sub-tasks generated from the user inputs) rather than simply responding to one-shot prompts.

Together, these features mark a departure from earlier forms of automation and code generation based on rules \cite{xu2008rule}, classical machine learning models \cite{TraditionalCodeGenration} or one-shot LLM calling \cite{DBLP:conf/pldi/polycoder22}. AI agentic programming represents a change toward intelligent systems that actively participate in the software development process. This enables new capabilities in intelligent code assistance \cite{github_copilot}, autonomous debugging and testing \cite{englhardt2024exploring, deng2024pentestgptllmempoweredautomaticpenetration}, automated code maintenance \cite{fan2023automated}, and potentially even self-improving software systems \cite{DBLP:journals/corr/abs-2405-15189, DBLP:journals/corr/abs-2310-02304}.

\subsubsection{Working example}

\begin{figure}[t!]
    \centering
    \includegraphics[width=0.8\linewidth]{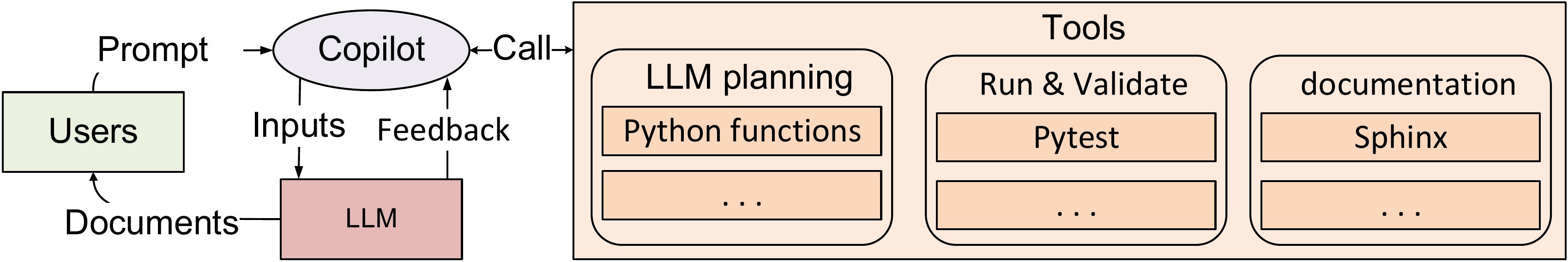}
    \caption{Agentic workflow for implementing a REST task.}
    \label{fig:agent_loop}
\end{figure}

As an example of AI agentic programming, consider a developer who tasks an AI coding agent with the following request: \textit{``Implement a REST API endpoint that returns the top 10 most frequently accessed URLs from a web server log file. Include unit tests and documentation.''} This task requires integrating multiple software components, including file parsing, frequency analysis, web API implementation, testing, and documentation.
A high-level view of an agentic loop for solving this task is depicted in Figure~\ref{fig:agent_loop}. Here, an LLM begins by analyzing the natural language task and planning a sequence of actions. It first produces a Python function to parse the log file and count URL frequencies using a dictionary or a data analysis library like \texttt{collections.Counter}. Next, it implements a REST API endpoint using a web framework such as \texttt{Flask}, exposing a route like \texttt{/top-urls} that returns the computed result in JSON format.
The LLM then writes unit tests and calls a Python interpreter to execute the generated Python script in the terminal and collects output to validate both the parsing logic and the API usage. It runs these tests using a tool like \texttt{pytest}, identifies failing cases, and refines the implementation. If a test fails due to a corner case (e.g., missing fields or malformed input), the LLM goes back to change the Python code, e.g., by adding input validation. It repeats this process - running tools, interpreting results, and modifying the code - until all tests pass. 
Finally, the LLM can generate documentation strings for each function and call an external tool like \texttt{pdoc} or \texttt{Sphinx} to produce human-readable API documentation. The process concludes when the agent validates that the tests pass, the API behaves as expected, and the documentation is complete.
This example illustrates the core features of agentic programming: autonomous planning, tool integration, iterative refinement, and goal-directed behavior. Unlike one-shot code generation, the agent interacts continuously with tools, learns from feedback, and adapts its actions to deliver a complete and functional software component.

\subsection{Historical Context and Motivations}

\begin{figure}[t!]
    \centering
    \includegraphics[width=0.9\linewidth]{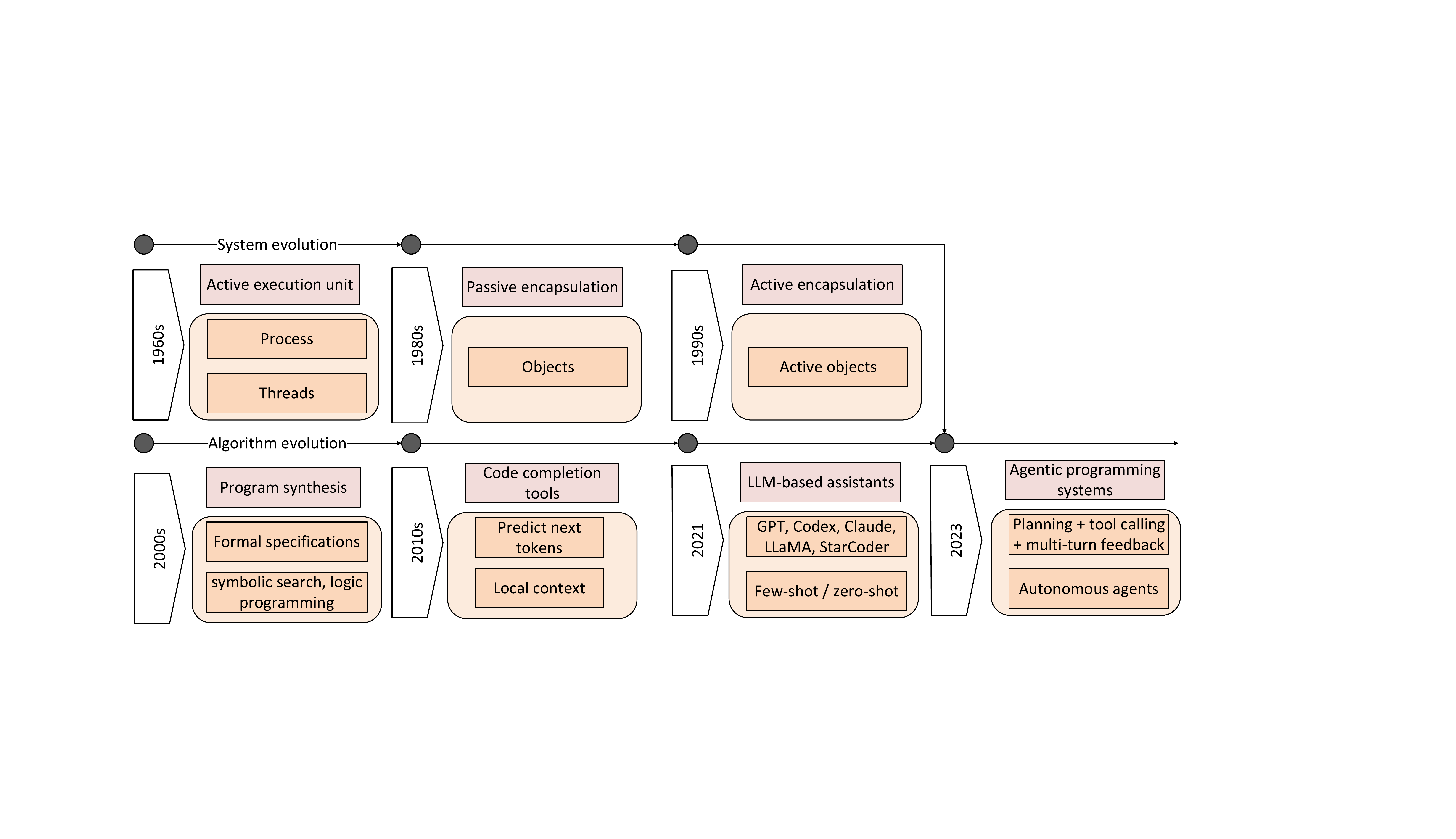}
    \caption{The evolution of coding agents from program synthesis, to code completion tools, to AI coding agents.}
    \label{fig:history}
\end{figure}

\begin{figure}[t!]
    \centering
    \includegraphics[width=0.9\linewidth]{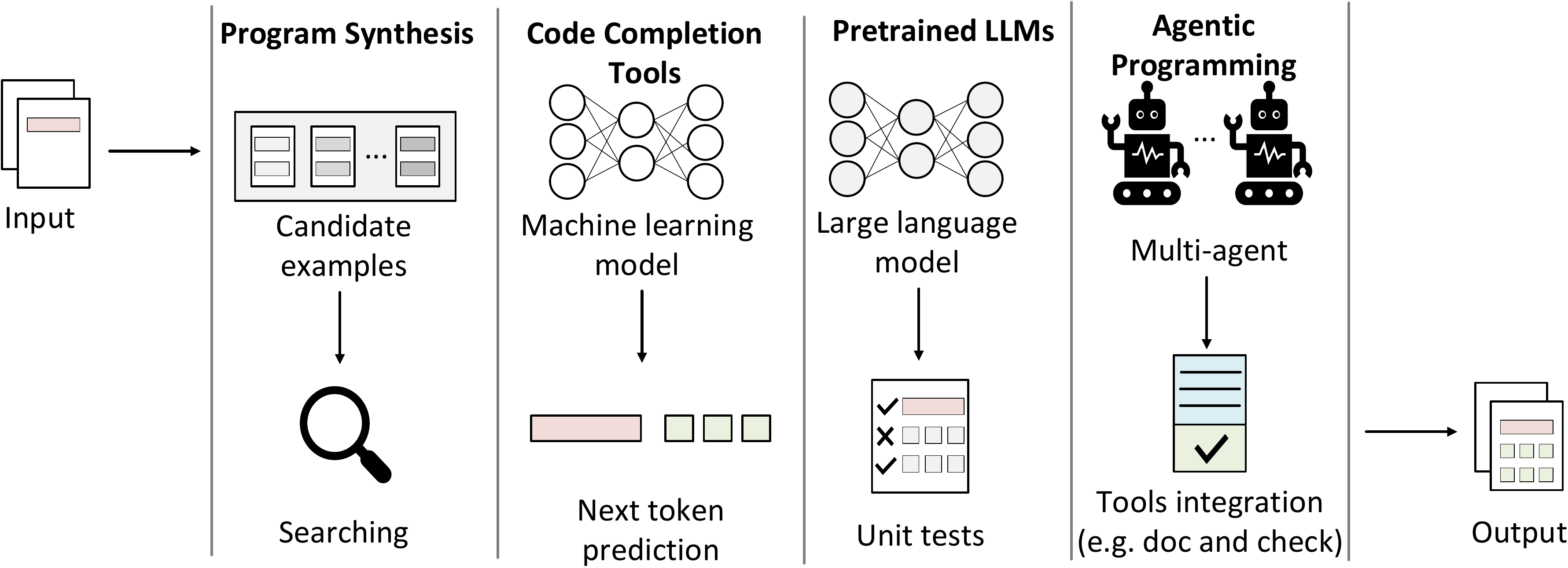}
    \caption{Evolution of code generation approaches.}
    \label{fig:workflow_code_ge}
\end{figure}

\subsubsection{A systems perspective}
As illustrated in Figure~\ref{fig:history}, the idea of automating software development has long been a goal in artificial intelligence and software engineering research~\cite{solar2008program}. 
In some respects, the rise of AI agents echoes the historical evolution of operating systems. Early operating systems introduced processes as the basic active execution units, capable of running independently. These evolved into threads, lightweight abstractions that enabled efficient concurrency and finer-grained scheduling~\cite{wollheim1984thread}. Later, with object-oriented design~\cite{riel1996object}, objects became the dominant unit of modularity - encapsulating state and behaviour, but remaining fundamentally passive. Active objects~\cite{nierstrasz1987active,nierstrasz1993regular} extended the object model by incorporating their own thread of control and asynchronous message passing, allowing them to act autonomously rather than merely responding to external calls—an early precursor to modern agents.

LLM-based AI agents can be seen as the next step in this trajectory. Like processes and threads, they are active entities, but unlike objects, they do not wait passively for instructions at every step. Instead, they can plan, act, and coordinate tasks proactively across diverse contexts. Their behavior is not restricted to fixed, handwritten policies but is driven by LLMs, enabling them to adapt dynamically, invoke external tools, and collaborate with both humans and other agents~\cite{li2024survey}. In this sense, agentic programming represents a new stage in the long-standing pursuit of making software development more autonomous, where code generation, debugging, and optimization are no longer fully hand-crafted but emerge through iterative interactions between humans, tools, and intelligent agents.

\subsubsection{Algorithm evolution of agents}
Figure~\ref{fig:workflow_code_ge} shows the evolution of algorithms. Early efforts in \textit{program synthesis} aimed to generate correct-by-construction programs from formal specifications~\cite{gulwani2017program}, while \textit{code completion tools} sought to improve developer productivity by predicting likely code snippets based on contexts~\cite{github_copilot, tabnine, amazon-q-developer}. These approaches, while impactful, typically relied on classical machine-learning models~\cite{TraditionalCodeGenration}, handcrafted ruless~\cite{xu2008rule}, or statistical techniques with limited generalizations~\cite{bruch2009learning}.

The advent of large-scale pre-trained LLMs such as Codexs~\cite{openai2021codex}, 
StarCoder~\cite{lozhkov2024starcoder},
LLaMas~\cite{meta2024llama3},
GPTs~\cite{openai2024gpt4o}, Claudes~\cite{claude2024}, 
Gemini~\cite{comanici2025gemini25pushingfrontier},
Qwen~\cite{hui2024qwen2},
and Deepseek~\cite{guo2025deepseek}, marked a major turning point. 
These models demonstrated strong zero-shot and few-shot capabilities in generating code~\cite{DBLP:conf/nips/Le0GSH22, peng2024perfcodegen}, translating between programming languages~\cite{DBLP:conf/emnlp/2020pymt5, DBLP:conf/mipro/GuzziKPGB24}, and answering complex programming questions with little or no task-specific fine-tuning~\cite{DBLP:journals/corr/abs-2203-07814, coignion2024performance}. Their ability to understand and generate natural language and code made them a good fit for software development tasks beyond basic code completion~\cite{husein2025large}, including documentation generation~\cite{dvivedi2024comparative}, test synthesis~\cite{jorgensen2024large}, and bug detection and fixing~\cite{englhardt2024exploring, deng2024pentestgptllmempoweredautomaticpenetration, kang2025explainable}.
As these models became more capable, a new opportunity emerged: using LLMs not just as passive code generators, but as autonomous agents that could reason about goals, invoke tools, and refine their outputs over multiple steps. This shift leads to the paradigm of AI agentic programming, where models operate as task-driven entities capable of planning, interacting with compilers and debuggers, and self-correcting based on feedback.

Several trends motivated this change. First, real-world software development often requires iterative problem solving, tool use, and adaptation, which single-step code generation cannot handle effectively~\cite{liu2024large, li2024survey}. Second, the rise of prompt engineering and structured prompting techniques (e.g., ReAct, chain-of-thought, scratchpads) enabled LLMs to reason more effectively over multiple steps~\cite{yao2023react, guo2025deepseek}. Third, the increasing availability of APIs, command-line tools, and language server protocols made it possible to integrate LLMs into full-stack development environments~\cite{anthropic_claude_cli, google_cloud_cli, anthropic_mcp_spec}.
These developments prompted a rethinking of how LLMs could be deployed-not just as smart coding assistants~\cite{github_copilot, tabnine}, but as semi-autonomous agents capable of carrying out software engineering tasks with minimal human supervision.

\subsection{Agency in AI Systems}

Agency is a foundational concept in the design of intelligent systems. At its core, an agent is an entity capable of perceiving its environment, reasoning about goals, and taking actions to influence outcomes. In the context of AI coding agents, \textit{agency} refers to a system's capacity to act autonomously, i.e., selecting actions based on internal objectives, external feedback, and learned knowledge.

Classical AI research has explored agency extensively in domains such as planning, robotics, and multi-agent systems~\cite{li2024survey, wu2024autogen}. Traditional agent models are typically characterized by four key attributes: \emph{reactivity}, the ability to respond to changes in the environment; \emph{proactivity}, the pursuit of long-term goals; \emph{social ability}, the capacity to communicate and coordinate with other agents or humans; and \emph{autonomy}, the ability to operate without direct human intervention.
In the context of AI agentic programming, these notions of agency are realized in new ways. An LLM-based system can interpret open-ended tasks expressed in natural language, plan sequences of development steps such as writing, testing, and debugging, and invoke or coordinate external tools like compilers, test runners, and linters. It can also adapt its actions in response to environmental feedback, such as compiler errors or test failures, while maintaining coherent state and reasoning across multiple iterations.
Unlike symbolic AI agents that rely on explicitly defined world models and search-based planning, LLM-based coding agents operate in a probabilistic, language-driven manner. Despite this difference, they increasingly exhibit behaviors aligned with classical definitions of agency, especially when augmented with memory, tool-use modules, and planning routines.

\subsection{Key Enablers of Agentic Programming}

Figure~\ref{fig:cagent} shows a representative system architecture of AI agentic programming. 
The emergence of AI agentic programming has been made possible by a combination of advances in language modeling, interaction frameworks, and software toolchains. Together, these enablers allow LLMs to move beyond static code generation toward goal-driven, interactive behavior. Below, we summarize the core technical factors that underpin this transition.

\subsubsection{Large language models}
\begin{table}[t]
\centering
\caption{Representative LLMs for coding tasks.}
\label{tab:llm-comparison}
\scriptsize
\begin{adjustbox}{width=\textwidth,center}
\begin{tabular}{p{1.9cm}rrlp{1.8cm}ccp{2.3cm}}
\toprule
\textbf{Model} & \textbf{Size (B)} & \textbf{Context Win.} & \textbf{Tool use} & \textbf{Provider (access)} & \textbf{Open Weight} & \textbf{MoE} & \textbf{Used in coding IDEs} \\
\midrule
GPT-5 & N/A & 1 M & \cmark & OpenAI    (API only) & \xmark & N/A & VS Code, Cursor, other IDEs \\
\rowcolor{gray!10} GPT-4 variants (o3, o4, etc.) & N/A & 128k & \cmark & OpenAI (API only) & \xmark & \xmark & VS Code, JetBrains, Cursor \\
Claude 4 Opus & $\sim$300 & 200k & \cmark & Anthropic (API only) & \xmark & \xmark & Cursor, Replit (chat) \\
\rowcolor{gray!10} Gemini 2.5 Pro & $\sim$200 & 1M  & \cmark & Google (API only) & \xmark & \cmark & Replit, Google Colab \\
Grok 4 & $\sim$1.7T & $\sim$128k & \cmark & xAI & \xmark & N/A & Not publicly integrated \\
\rowcolor{gray!10} DeepSeek R1-0528 & 671 (act. 37) & 160k & \cmark & DeepSeek (API + weights) & \cmark & \cmark & Emacs, VS Code (via extension) \\
Kimi K2 & 1000 (act. 32) & 128k & Limited & Moonshot AI (API) & \cmark & N/A & Custom plugin support \\
\rowcolor{gray!10} Qwen3-235B-A22B & 235 (act. 22) & 128k & Limited & Alibaba & \cmark & \cmark & Alibaba Cloud IDE \\
Qwen3-Coder-480B-A35B-Instruct & 480 (act. 35) & 256k &  & Alibaba & \cmark & \cmark & Alibaba Cloud IDE \\
% LLama 3.3 & 405 & 128k & Limited & Meta (weights available) & \cmark & \xmark & VS Code (open source extensions) \\
% StarCoder2 & 15 & 128k & Limited & Hugging Face (weights) & \cmark & \xmark & JetBrains, VS Code (via extensions) \\
 \rowcolor{gray!10} Solar-Pro & 72 & 128k & \cmark & Upstage (weights on HF) & \cmark & \xmark & VS Code (via third-party) \\
% Mistral 7B & 7 & 32k & Limited & Mistral (HF weights) & \cmark & \xmark & VS Code (open source support) \\
% Mistral Large & $\sim$35 & 32k & \cmark & Mistral (API only) & \xmark & \xmark & Le Chat, VS Code (via API) \\
Openhands-LM-32B-v0.1 & 32 & 128k & \cmark & OpenHands & \cmark & \xmark & VS Code (via extension) \\
\rowcolor{gray!10} Devstral-Medium & N/A & 128k & \cmark & Mistral (API only) & \xmark & N/A & VS Code (via API) \\
Devstral-Small & 24 & 128k & \cmark & Mistral (API only) & \cmark & \xmark & VS Code (via API) \\

\bottomrule
\end{tabular}
\end{adjustbox}
\end{table}

LLMs trained on massive corpora of code and natural language form the foundation of modern agentic programming systems. These models, as represented by GPT-5 \cite{openai2025gpt5introducing}, Claude \cite{claude2024}, DeepSeek \cite{guo2025deepseek}, and Gemini \cite{gemini2023}, serve as the core reasoning engines, powering code generation, task planning, debugging, documentation, and natural language interaction. Their ability to understand and execute complex instructions makes them central to the design of agentic workflows.
Modern LLMs can generate syntactically correct and semantically meaningful code, answer development-related queries, and engage in multi-turn conversations with minimal task-specific fine-tuning. Many of these models leverage few-shot, zero-shot, and in-context learning capabilities, allowing them to generalize across programming languages, frameworks, and task domains. This flexibility enables developers to use the same underlying model for a wide range of software engineering tasks, from scaffolding and unit test generation to bug repair and performance tuning. In addition to general-purpose models, some LLMs like Grok \cite{grok4} and Calude Opus \cite{anthropic_opus4} are increasingly optimized for coding tasks through specialized instruction tuning, extended context length, tool use capabilities, and integration with retrieval-based systems. These enhancements make them suitable for multi-turn reasoning, code synthesis grounded in external context, and tool-augmented workflows.
Table~\ref{tab:llm-comparison} provides a comparative overview of some of the state-of-the-art LLMs used in code-related tasks. The table compares their key attributes. As the capabilities of LLMs continue to evolve, selecting and fine-tuning the appropriate foundation model for coding tasks becomes a critical design choice in building reliable, efficient, and adaptive agentic systems.

\subsubsection{Prompt engineering and reasoning strategies}
Effective agentic behavior often requires structured prompting techniques to guide LLMs through multi-step reasoning and tool use. Rather than relying on a single input-output exchange, these methods provide scaffolding that helps models break tasks into manageable steps and maintain coherence across longer interactions. For instance, chain of thought prompting~\cite{wei2022chain} encourages explicit reasoning traces, making intermediate steps visible and improving problem-solving accuracy. ReAct (reasoning and acting) interleaves reasoning with concrete actions~\cite{yao2023react}, such as tool calls or environment interactions, enabling agents to both deliberate and act in context. Scratchpad prompting~\cite{abbe2024far} provides a working memory space where partial results, hypotheses, or plans can be written down and refined, supporting iterative refinement. Modular prompting~\cite{sun-etal-2023-multitask}, on the other hand, separates tasks into distinct functional roles—such as planner, executor, and verifier—so that the model can coordinate across specialized subtasks.
Together, these techniques allow agents to decompose complex problems, retain intermediate states, and revise their behavior in light of new evidence. They also increase transparency by exposing the reasoning process and provide greater controllability by constraining how models structure their outputs. In practice, structured prompting forms the backbone of many agentic systems, enabling LLMs to move beyond ad hoc responses and toward more reliable, interpretable, and goal-directed behavior.

\subsubsection{Tool use and API integration}
\begin{table}[t]
\centering
\caption{Examples of tools supported by GitHub Copilot agent.}
\label{tab:copilot-tools}
\scriptsize
\begin{adjustbox}{width=0.6\textwidth,center}
\begin{tabular}{ll}
\toprule
\textbf{Tool Type} & \textbf{Examples} \\
\midrule
Compiler & \texttt{gcc}~\cite{GCC}, \texttt{clang}~\cite{Clang}, \texttt{javac}~\cite{Javac}, \texttt{tsc}~\cite{TSC} \\
% \hline
\rowcolor{gray!10} Debugger & \texttt{gdb}~\cite{GDB}, \texttt{lldb}~\cite{LLDB}, \texttt{pdb}~\cite{PDB} \\
% \hline
Test Framework & \texttt{pytest}~\cite{Pytest}, \texttt{unittest}~\cite{Unittest}, \texttt{Jest}~\cite{Jest}, \texttt{Mocha}~\cite{Mocha} \\
% \hline
\rowcolor{gray!10} Linter & \texttt{eslint}~\cite{ESLint}, \texttt{flake8}~\cite{Flake8}, \texttt{black}~\cite{Black}, \texttt{prettier}~\cite{Prettier} \\
% \hline
Version Control & \texttt{git}~\cite{Git} \\
% \hline
\rowcolor{gray!10} Build System & \texttt{make}~\cite{Make}, \texttt{cmake}~\cite{CMake}, \texttt{npm}~\cite{NPM}, \texttt{maven}~\cite{Maven} \\
% \hline
Package Manager & \texttt{pip}~\cite{Pip} \texttt{yarn}~\cite{Yarn}, \texttt{cargo}~\cite{Cargo}\\
% \hline
\rowcolor{gray!10} Language Server & \texttt{pyright}~\cite{Pyright}, \texttt{tsserver}~\cite{TSServer} \\
\bottomrule
\end{tabular}
\end{adjustbox}
\end{table}

\begin{figure}[t]
\centering
\begin{lstlisting}[language=JavaScript,basicstyle=\scriptsize\ttfamily,caption={An example of the OpenAI tool schema},  numbers=none,
   frame=lines, stringstyle=\color{green!50!black}, label={lst:compiler-tool-schema}]
import OpenAI from "openai";
const client = new OpenAI();
const tools = [
{
  "type": "function",
  "function": {
    "name": "compile_code",
    "description": "...",
    "parameters": {
      "type": "object",
      "properties": {
        "language": {
          "type": "string",
          "enum": ["c", "cpp"],
          "description": "Programming language to compile."
        },
        "source": {
          "type": "string",
          "description": "Source code to compile."
        },
        "flags": {
          "type": "array",
          "items": { "type": "string" },
          "description": "Compiler flags."
        }
      },
      "required": ["language", "source"],
      "additionalProperties": false
    }
  }
},
];
\end{lstlisting}
\label{fig:compiler-tool-schema}
\end{figure}

\begin{table}[t]
\centering
\caption{Example interfaces between LLMs and tools.}
\label{tab:llm-tool-interfaces}
\scriptsize
\begin{adjustbox}{width=0.95\textwidth,center}
\begin{tabular}{lll}
\toprule
\textbf{Interface Type} & \textbf{Description} & \textbf{Example} \\
\midrule
Natural Language 
& LLM sends plain text instructions; the tool parses heuristically. 
& ``Run my tests and show errors'' \\
\rowcolor{gray!10}
Command Line 
& Tools invoked with shell commands; I/O as text streams. 
& \texttt{gcc main.c -O2} \\
Language Server Protocol 
& JSON-RPC protocol exposing AST, symbols, diagnostics. 
& VS Code using LSP for Python \\
\rowcolor{gray!10}
REST / gRPC APIs 
& Tools exposed as network services with structured request/response. 
& GitHub Actions REST API \\
Structured Schema (JSON) 
& Actions, parameters, outputs defined in machine-readable schema. 
& OpenAI function calling JSON schema \\
\rowcolor{gray!10}
Intermediate Representation 
& LLM interacts with compiler/runtime IR or AST. 
& LLVM IR for code optimization \\
Event Stream / Logs 
& Tools output execution traces or state changes as structured streams. 
& Debug logs from pytest \\
\rowcolor{gray!10}
Framework-based Adapters 
& Middleware unifies heterogeneous tool interfaces. 
& LangChain, Model Context Protocol \\
\bottomrule
\end{tabular}
\end{adjustbox}
\end{table}

Agentic systems rely heavily on external tools, such as compilers, debuggers, test frameworks, linters, and version control systems, to validate and refine generated code. These tools provide the concrete signals needed to check correctness, enforce coding standards, and ensure that outputs remain consistent with project requirements. For example, Table~\ref{tab:copilot-tools} lists a subset of the tools currently supported by the GitHub Copilot Agent~\cite{github_copilot}, covering compilation, testing, and version control.
Integration with external tools can take multiple forms, including command line interfaces, language server protocols (LSP), and RESTful APIs. Increasingly, LLMs interact with tools through structured Python or JavaScript interfaces that specify the available actions, input parameters, and expected outputs in a machine-readable format. For example, Listing~\ref{lst:compiler-tool-schema} shows how a compiler can be exposed as a tool to extend an LLM’s capabilities. This structured approach reduces ambiguity, grounds commands in the correct syntax, and makes it easier for the model to call external tools safely and consistently. By interpreting the schema, an LLM can generate well-formed commands, parse structured responses, and adjust its behavior in a predictable way. Table~\ref{tab:llm-tool-interfaces} summarizes the main types of interfaces through which LLMs interact with external tools. These range from free-form natural language instructions to highly structured schemas and domain-specific protocols.

\subsubsection{State and context management}

LLMs operate under fixed context windows, limiting their ability to reason over long histories. Agentic systems therefore incorporate external memory mechanisms to store plans, results, tool outputs, and partial progress. This memory can take the form of vector stores, scratchpads, or structured logs, allowing the agent to recall relevant information across multiple steps and maintain coherence over long-running tasks.
Table~\ref{tab:context-management} compares the context management strategies of mainstream AI coding agents, revealing substantial differences in context size and memory persistence. Tools like GitHub Copilot~\cite{github_copilot} currently do not utilize persistent memory, instead using transient methods such as sliding windows or dynamic token budgeting. In contrast, agents like SWE-agent~\cite{yang2024swe}, Devika~\cite{devika}, and OpenDevin~\cite{OpenDevin} employ persistent storage, often via vector databases or structured stores, to support long-term recall of plans, tool outputs, and project history. Some, such as Cursor IDE~\cite{cursor_editor} and Continue.dev~\cite{continue}, use embedding-based search to retrieve semantically relevant content, while others summarize prior actions to stay within the available context window.
These differences reflect a clear trade-off: smaller context windows typically rely on lightweight retrieval or summarization, whereas larger windows with persistent memory enable richer state tracking but add storage and retrieval overhead.

\subsubsection{Feedback loops and self-improvement}
Agentic programming leverages feedback to refine outputs iteratively. Agents may rerun failed tests, revise prompts based on compiler errors, or reflect on past failures to improve future behavior. For instance, compiler errors may trigger targeted code edits, test failures may prompt iterative debugging, and linter warnings may guide stylistic refinements. Some systems incorporate explicit planning, retry mechanisms, or even gradient-based updates~\cite{huang2024understanding} through fine-tuning or reinforcement learning. This closed-loop design supports robustness and adaptability in complex programming tasks.

\subsection{Comparison to Related Paradigms}

AI agentic programming represents a distinct paradigm that builds upon but fundamentally differs from existing paradigms that have shaped the landscape of automated software development.

\subsubsection{Program synthesis}
Program synthesis has been a foundational approach to automated code generation, traditionally divided into two types: \emph{deductive synthesis} uses formal specifications to generate provably correct programs, while \emph{inductive synthesis} learns from input-output examples with symbolic search and logic programming techniques to infer program logic~\cite{gulwani2017program, jain2022jigsaw}. Classical synthesis systems like sketching~\cite{solar2008program} and more recent neural approaches like RobustFill~\cite{devlin2017robustfill} specialize in generating targeted code snippets that satisfy precise specifications.
However, classical program synthesis focuses on single-function generation from formal specifications (due to the scaling challenge of code sysnthesis~\cite{gulwani2017program}) and typically operates in a one-shot generation mode~\cite{devlin2017robustfill}, whereas agentic programming handles multi-step workflows (e.g., planning, tool use, and iterative refinement) and engages in continuous interaction with development environments \cite{bouzenia2024repairagent}.

\begin{table}[t]
\centering
\scriptsize
\caption{Context management mechanisms supported by mainstream AI coding agents.}
\label{tab:context-management}
\begin{adjustbox}{width=\textwidth,center}
\begin{tabular}{lrccl}
\toprule
\textbf{Agent} & \textbf{Underlying Model} & \textbf{Context Window (default)} & \textbf{Persistent Memory} & \textbf{Context Management Mechanism} \\
\midrule
GitHub Copilot & GPT-4 (o-series) & 16k & \xmark & Sliding window over active buffer \\
\rowcolor{gray!10} Codeium & GPT-4 / Claude 3.5 & 32k & \xmark & Dynamic token budgeting based on file proximity and edit history \\
Cursor IDE & Claude 3.5 Sonnet / GPT-4 & 128k & \cmark & Semantic search over project history \\
\rowcolor{gray!10} SWE-agent & GPT-4 & 16k & \cmark & Vector DB retrieval for tool outputs and plan state \\
Devika & GPT-4 / Open LLM & 32k & \cmark & Structured memory via SQLite and embeddings \\
\rowcolor{gray!10} AutoDev & GPT-4 & 16k & \cmark & Summarization of prior actions and tool logs \\
Continue.dev & GPT-4 / Claude & 32k & \xmark & Embedding-based local recall over recent edits \\
\rowcolor{gray!10} OpenDevin & GPT-4 / Claude / Mixtral & 32k & \cmark & RAG over command history, plans, and intermediate outputs \\

\bottomrule
\end{tabular}
\end{adjustbox}
\end{table}

\subsubsection{Code completion tools}

Code completion, as one of the most commercially successful applications of AI in programming, excels at context-aware code suggestion, leveraging large-scale pre-training on code repositories to predict next tokens of partially written code \cite{DBLP:conf/iclr/codegen23, DBLP:journals/corr/abs-2203-07814, husein2025large}. Advanced completion tools can suggest entire functions, classes, or small modules based on comments, function signatures, and surrounding context, with tools like GitHub Copilot \cite{github_copilot}, TabNine \cite{tabnine}, and Amazon Q Developer \cite{amazon-q-developer} achieving widespread adoption.
These code completion tools often operate as reactive assistants that respond to developer input, while agentic programming systems demonstrate proactive behavior and autonomous planning. Furthermore, agentic programming extends beyond code generation to encompass testing, debugging, deployment, and maintenance activities that completion tools typically do not address~\cite{yang2024swe, bouzenia2024repairagent}.

\subsubsection{DevOps automation}

DevOps automation focuses on streamlining software delivery pipelines through Infrastructure as Code (IaC), Continuous Integration/Continuous Deployment (CI/CD), and automated testing frameworks~\cite{humble2010continuous}. Tools like Jenkins \cite{jenkins}, GitLab CI \cite{gitlab_ci}, and modern platforms like GitHub Actions \cite{github_actions} automate repetitive deployment tasks, testing workflows, and infrastructure management.
While both paradigms emphasize automation, DevOps automation primarily handles pre-defined workflows and infrastructure management, whereas agentic programming focuses on adaptive problem-solving and creative solution generation. Additionally, agentic programming can potentially orchestrate and improve DevOps processes themselves, representing a higher-order form of automation \cite{kambala2024intelligent}.

\subsubsection{Automated machine learning and automated development}

Automated Machine Learning (AutoML) represents a successful paradigm for democratizing AI model development through automation of model selection, hyperparameter tuning, and feature engineering~\cite{he2021automl}. Platforms like Google Cloud AutoML \cite{google_automl}, Amazon SageMaker Autopilot \cite{sagemaker_autopilot}, and open-source frameworks like Auto-sklearn automate the traditional machine learning pipeline from data preprocessing to model deployment~\cite{feurer2015efficient}.
However, AutoML focuses on statistical model optimization within well-defined machine learning workflows and operates with structured data and standardized evaluation metrics, while agentic programming tackles more general software development challenges with creative problem-solving and multi-modal reasoning. 

\subsubsection{Multi-Agent systems and human-AI collaboration}

Traditional multi-agent systems in software development typically involve specialized agent roles working within predefined coordination protocols~\cite{jennings2000agent}. These systems often feature separate agents for requirements analysis, code generation, testing, and documentation, coordinating through structured communication interfaces. Recent advances in LLM-based multi-agent programming have demonstrated the potential for more sophisticated collaboration, with systems like MetaGPT showing how multiple AI agents can simulate software development teams~\cite{hong2023metagpt}.
AI agentic programming can be viewed as an evolution of multi-agent systems that incorporates human-in-the-loop collaboration and dynamic role adaptation. Unlike traditional multi-agent systems with fixed agent roles and rigid communication protocols, agentic programming systems demonstrate fluid role assignment and context-adaptive behavior. Moreover, the integration of tool use, environmental interaction, and persistent memory distinguishes modern agentic programming from earlier multi-agent approaches. Contemporary agentic systems like AutoGen~\cite{wu2024autogen} and CrewAI~\cite{duan2024exploration} enable agents to directly interact with development tools, maintain context across extended sessions, and learn from past interactions~\cite{li2024survey}. This represents a significant advancement over traditional multi-agent systems that typically operated in more constrained, simulation-based environments.

\subsubsection{Comparison with robotics and reinforcement learning agents}  
Robotic agents typically interact with the physical world through sensors and actuators. Their perception, control, and planning modules are tightly coupled with real-time feedback and often require safety guarantees. Reinforcement learning (RL) agents~\cite{sun2024llm}, by contrast, learn behaviors by maximizing cumulative reward through trial and error, often in simulated environments. These agents explore large state-action spaces and acquire policies over time.
Agentic programming systems share features with both paradigms. Like robotic agents, coding agents must coordinate perception, such as task understanding, with actions such as code edits or tool invocations, all within an environment shaped by constraints and feedback. Like RL agents, they benefit from feedback loops, for example, test results or compiler outputs, and may employ exploration, retry strategies, or even reward-guided behavior.
At the same time, coding agents differ in multiple ways. They operate in a symbolic, tool-rich environment where actions are language-based and environments, such as codebases, APIs, and test harnesses, are highly structured. Success requires reasoning not only about immediate feedback but also about abstract software goals, dependencies, and long-term coherence across multiple steps. This makes agency in programming uniquely challenging and distinct from both the physical world and simulated agents.

\section{Survey Methodology}
\label{sec:methodology}

\begin{figure}[!t]
\centering
\begin{minipage}{0.62\textwidth}
\includegraphics[width=\columnwidth]{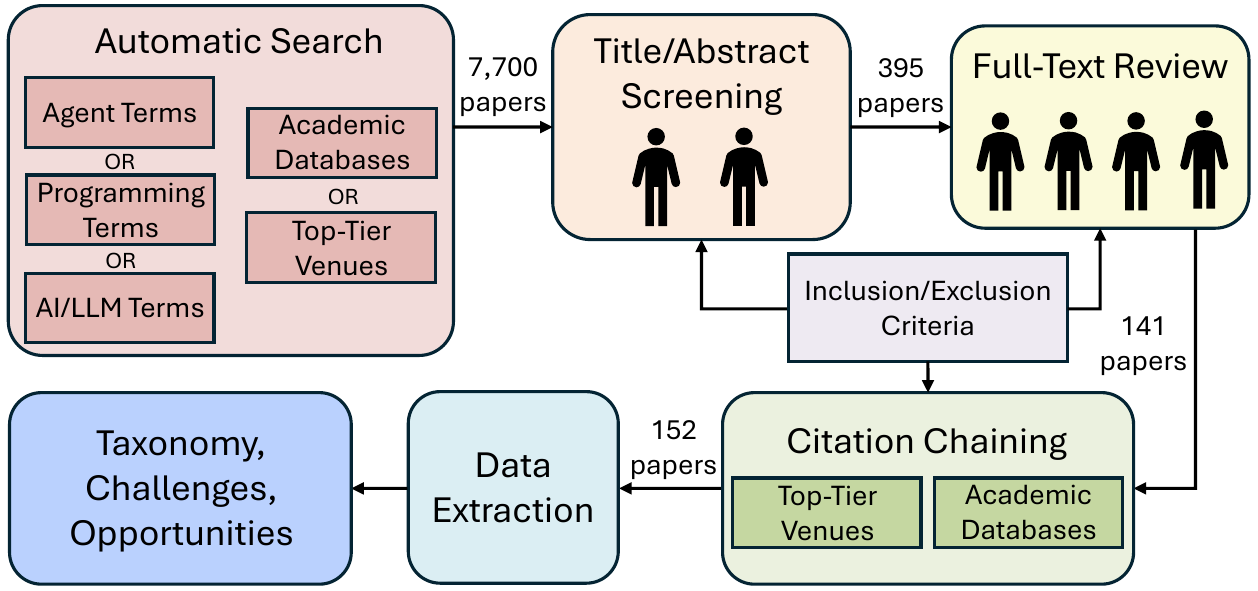}
% \vspace{-0.7cm}
  \caption{Survey methodology for academic research.}
 \label{fig:methodology}
  \end{minipage}
~\hspace{-0.3cm}
\begin{minipage}{0.32\textwidth}
\centering
\begin{tikzpicture}[scale=0.7]
\tikzstyle{every node}=[font=\large]
\definecolor{beaublue}{rgb}{0.74, 0.83, 0.9}
\definecolor{applegreen}{rgb}{0.55, 0.71, 0.0}
\definecolor{dollarbill}{rgb}{0.52, 0.73, 0.4}
\definecolor{ao(english)}{rgb}{0.0, 0.5, 0.0}
\definecolor{amaranth}{rgb}{0.9, 0.17, 0.31}
\definecolor{cerublue}{rgb}{0.16, 0.32, 0.75}
\definecolor{frenchblue}{rgb}{0.0, 0.45, 0.73}
\definecolor{iceberg}{rgb}{0.44, 0.65, 0.82}
\definecolor{myred}{rgb}{0.898,0.725,0.710}
\definecolor{myorange}{rgb}{0.988,0.922,0.867}
\definecolor{myyellow}{rgb}{0.984,0.980,0.855}
\definecolor{mygreen}{rgb}{0.922,0.945,0.875}
  
  % % Draw the second slice (4%)
  % \pie[color={blue},rotate=342]{4/}
  % 0.95×360=342
  
  % % Draw the third slice (1%)
  % \pie[color={green},rotate=356.4]{1/}
  % 0.99×360=356.4
    \pie[
        /tikz/every pin/.style={align=center},
        % text=legend,
        text=pin,
        % before number=\phantom,
        % after number=,
        pin distance = 10,
        rotate=260,
        explode=0, 
        color={mygreen, myorange, myred, myyellow},
        ]
        {
        5/\texttt{2022} (7),
        22/\texttt{2023} (33),
        53/\texttt{2024} (81),
        20/\texttt{2025} (30)
        }

    \end{tikzpicture}
% ~\vspace{-0.13cm}
\caption{Distribution of academic papers.}
\label{fig:pie_year}
  \end{minipage}
 \end{figure}

This survey follows a widely-used systematic literature review (SLR) methodology~\cite{keele2007guidelines, DBLP:journals/corr/Gong24Deep, hou2024largelanguagemodelssoftware, DBLP:journals/tse/WangHGGZFSLZN23, zhang2024systematic, qudus2025fact} to provide comprehensive coverage of AI agentic programming research, as illustrated in Figure~\ref{fig:methodology}.

\subsection{Search Strategy}

We conducted automatic searches across multiple academic databases, including Google Scholar, ACM Digital Library, IEEE Xplore, SpringerLink, and arXiv.org. We also examined proceedings from top-tier venues (FSE, ICSE, ASE, ICML, NeurIPS, AAAI, etc.). As this fast-evolving field is largely dominated by industry, we also pay attention to open-source and the industry releases of relevant results and tools. 

Our search string combined the following term clusters using Boolean operators:

\keybox{

\begin{itemize}
    \item \textbf{Agent terms}: ``AI agent'' OR ``agentic'' OR ``autonomous agent'' OR ``coding agent'' OR ``software agent'' OR ``intelligent agent'' OR ``task agent'' OR ``LLM agent''
    
    \item \textbf{Programming terms}: ``programming'' OR ``coding'' OR ``software development'' OR ``code generation'' OR ``software engineering'' OR ``developer'' OR ``autonomous coding'' OR ``software automation''

    \item \textbf{AI/LLM terms}: ``large language model'' OR ``LLM'' OR ``language model'' OR ``foundation model'' OR ``AI model'' OR ``neural code generation''
\end{itemize}

}

\subsection{Study Selection}
After initial retrieval, we followed a three-stage study selection process for academic papers: (1) title and abstract screening by two independent researchers, (2) full-text review with disagreement resolution through discussion, and (3) backward and forward citation chaining to identify additional relevant studies. During the selection process, we used the following criteria:

\textbf{Inclusion criteria} - studies were included if they met all of the following:
\begin{enumerate}
    \item Focus on AI systems for software development with autonomous/semi-autonomous behavior
    \item Demonstrate agentic behaviors: planning, tool use, iterative refinement, or adaptive decision-making
    \item Present novel techniques, architectures, evaluations, or comprehensive analysis
    \item Include experimental evaluation, case studies, or substantial implementation details
    \item Written in English with accessible full text
\end{enumerate}

\textbf{Exclusion criteria} - studies were excluded if they:
\begin{enumerate}
    \item Focused solely on traditional code completion without agentic behavior
    \item Addressed non-programming domains (robotics, game playing, etc.)
\end{enumerate}

\subsection{Results}

Our systematic search yielded:
\begin{itemize}
    \item \textbf{Initial retrieval}: 7,700 papers from database searches
    \item \textbf{Title/abstract screening}: 395 papers selected for full-text review
    \item \textbf{Full-text review}: 141 met all criteria.
    \item \textbf{Final corpus}: 152 papers included after full-text evaluation and citation chaining
    \item \textbf{Software tools and industry products}: we also study a wide range of state-of-the-art AI coding agents and LLMs like GitHub Copilot Agents, GPT, Gemini, Deepseek, Qwen and Claude Opus 4. 
\end{itemize}

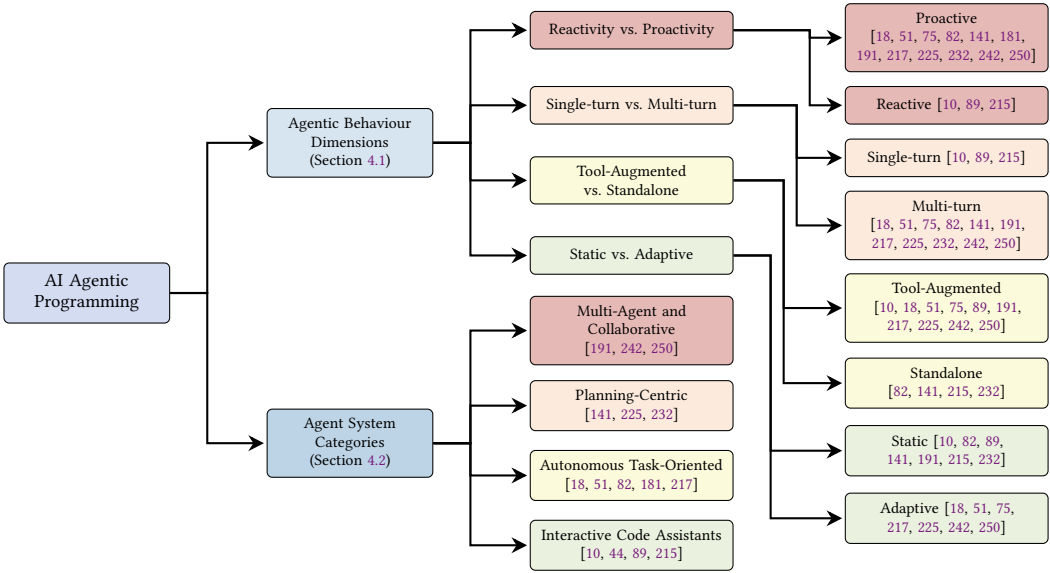
\begin{figure}[t]
    \centering
    \begin{tikzpicture}[
    every node/.style={draw, rounded corners=2pt, minimum height=0.6cm, text width=1.8cm, align=center, font=\tiny},
    root/.style={fill=cerublue!20, font=\scriptsize, text width=2cm, minimum height=0.8cm},
    dimension/.style={fill=beaublue!60, font=\tiny, text width=2cm, minimum height=0.7cm},
    dimension2/.style={fill=beaublue!100, font=\tiny, text width=2cm, minimum height=0.7cm},
    category/.style={fill=applegreen!60, font=\tiny, text width=2.5cm, minimum height=0.6cm},
    system/.style={fill=cadmiumorange!40, font=\tiny, text width=2.5cm, minimum height=0.5cm},
    category1/.style={fill=myred, font=\tiny, text width=2.5cm, minimum height=0.5cm},
    category2/.style={fill=myorange, font=\tiny, text width=2.5cm, minimum height=0.5cm},
    category3/.style={fill=myyellow, font=\tiny, text width=2.5cm, minimum height=0.5cm},
    category4/.style={fill=mygreen, font=\tiny, text width=2.5cm, minimum height=0.5cm}
]

% Root node
\node[root] (root) {AI Agentic Programming};

% First level - Agentic Behaviour Dimensions
\node[dimension, right=\leveldist of root, xshift=0.3cm, yshift=\yshiftone] (behav) {Agentic Behaviour\\Dimensions\\(Section \ref{subsec:4.1})};

% First level - Agent System Categories  
\node[dimension2, right=\leveldist of root, xshift=0.3cm, yshift=-\yshiftone] (syscat) {Agent System\\Categories\\(Section \ref{subsec:4.2})};

% Second level under Agentic Behaviour Dimensions
\node[category1, right=\leveldist of behav, xshift=0.3cm, yshift=\yshifttwo] (react) {Reactivity vs. Proactivity};
\node[category2, right=\leveldist of behav, xshift=0.3cm, yshift=\yshiftthree] (turn) {Single-turn vs. Multi-turn};
\node[category3, right=\leveldist of behav, xshift=0.3cm, yshift=-\yshiftthree] (tool) {Tool-Augmented vs. Standalone};
\node[category4, right=\leveldist of behav, xshift=0.3cm, yshift=-\yshifttwo] (static) {Static vs. Adaptive};

% Second level under Agent System Categories
\node[category1, right=\leveldist of syscat, xshift=0.3cm, yshift=\yshifttwo] (multiagent) {Multi-Agent and Collaborative\\\cite{yang2024swe, qian2024chatdevcommunicativeagentssoftware, zhang2024autocoderover}};
\node[category2, right=\leveldist of syscat, xshift=0.3cm, yshift=\yshiftthree] (planning) {Planning-Centric\\\cite{wang2023voyager, li2023camel, wen2024codeplan}};
\node[category3, right=\leveldist of syscat, xshift=0.3cm, yshift=-\yshiftthree+2] (autonomous) {Autonomous Task-Oriented\\\cite{openai2025gpt5introducing, anthropic_opus4, Jules, team2025kimi, comanici2025gemini25pushingfrontier}};
\node[category4, right=\leveldist of syscat, xshift=0.3cm, yshift=-\yshifttwo+4] (interactive) {Interactive Code Assistants\\\cite{github_copilot, cursor, amazon-q-developer, tabnine}};

% Third level under Reactivity vs. Proactivity - positioned side by side with more spacing
\node[category1, right=\leveldist of react, xshift=0.5cm, yshift=-0.1cm] (react1) {Proactive \cite{openai2025gpt5introducing, anthropic_opus4, Jules, devgpt2024, team2025kimi, comanici2025gemini25pushingfrontier, wang2023voyager, li2023camel, wen2024codeplan, qian2024chatdevcommunicativeagentssoftware, zhang2024autocoderover, yang2024swe}};
\node[category1, right=\leveldist of react, xshift=0.5cm, yshift=-1cm] (react2) {Reactive \cite{github_copilot, tabnine, cursor}};

% Third level under Single-turn vs. Multi-turn - positioned side by side with more spacing
\node[category2, right=\leveldist of turn, xshift=0.5cm, yshift=-0.7cm] (turn1) {Single-turn \cite{github_copilot, tabnine, cursor}};
\node[category2, right=\leveldist of turn, xshift=0.5cm, yshift=-1.6cm] (turn2) {Multi-turn \cite{anthropic_opus4, Jules, devgpt2024, team2025kimi, comanici2025gemini25pushingfrontier, wang2023voyager, li2023camel, wen2024codeplan, qian2024chatdevcommunicativeagentssoftware, zhang2024autocoderover, yang2024swe}};

% Third level under Tool-Augmented vs. Standalone - positioned side by side with more spacing
\node[category3, right=\leveldist of tool, xshift=0.5cm, yshift=-1.7cm] (tool1) {Tool-Augmented \cite{github_copilot, cursor, anthropic_opus4, Jules, devgpt2024, team2025kimi, wang2023voyager, qian2024chatdevcommunicativeagentssoftware, zhang2024autocoderover, yang2024swe}};
\node[category3, right=\leveldist of tool, xshift=0.5cm, yshift=-2.7cm] (tool2) {Standalone \cite{tabnine, comanici2025gemini25pushingfrontier, li2023camel, wen2024codeplan}};

% Third level under Static vs. Adaptive - positioned side by side with more spacing
\node[category4, right=\leveldist of static, xshift=0.5cm, yshift=-2.6cm] (static1) {Static \cite{github_copilot, tabnine, cursor, comanici2025gemini25pushingfrontier, li2023camel, wen2024codeplan, qian2024chatdevcommunicativeagentssoftware}};
\node[category4, right=\leveldist of static, xshift=0.5cm, yshift=-3.5cm] (static2) {Adaptive \cite{anthropic_opus4, Jules, devgpt2024, team2025kimi, wang2023voyager, zhang2024autocoderover, yang2024swe}};

% Draw two-bend (three-segment) right-angled lines from root to first level
\draw[arrowstyle] (root.east) -- ++(\bend,0) |- (behav.west);
\draw[arrowstyle] (root.east) -- ++(\bend,0) |- (syscat.west);

% Draw two-bend lines from first level to second level
\draw[arrowstyle] (behav.east) -- ++(\bend,0) |- (react.west);
\draw[arrowstyle] (behav.east) -- ++(\bend,0) |- (turn.west);
\draw[arrowstyle] (behav.east) -- ++(\bend,0) |- (tool.west);
\draw[arrowstyle] (behav.east) -- ++(\bend,0) |- (static.west);

\draw[arrowstyle] (syscat.east) -- ++(\bend,0) |- (multiagent.west);
\draw[arrowstyle] (syscat.east) -- ++(\bend,0) |- (planning.west);
\draw[arrowstyle] (syscat.east) -- ++(\bend,0) |- (autonomous.west);
\draw[arrowstyle] (syscat.east) -- ++(\bend,0) |- (interactive.west);

% Draw two-bend lines from second level to third level
\draw[arrowstyle] (react.east) -- ++(\bend+5+5+5,0) |- (react1.west);
\draw[arrowstyle] (react.east) -- ++(\bend+5+5+5,0) |- (react2.west);

\draw[arrowstyle] (turn.east) -- ++(\bend+5+5,0) |- (turn1.west);
\draw[arrowstyle] (turn.east) -- ++(\bend+5+5,0) |- (turn2.west);

\draw[arrowstyle] (tool.east) -- ++(\bend+5,0) |- (tool1.west);
\draw[arrowstyle] (tool.east) -- ++(\bend+5,0) |- (tool2.west);

\draw[arrowstyle] (static.east) -- ++(\bend,0) |- (static1.west);
\draw[arrowstyle] (static.east) -- ++(\bend,0) |- (static2.west);

\end{tikzpicture}
    \caption{Taxonomy of AI agentic programming systems.}
    % \FIXME{Need to update \cite{amazon-q-developer} for Sec 4.1 taxonomy}
    \label{fig:taxonomy-overview}
    \end{figure}

Among the 152 academic references published from 2022 to 2025 (excluding tool descriptions and websites), 5\% appeared in 2022, 22\% in 2023, 53\% in 2024, and 20\% in 2025, as shown in Figure~\ref{fig:pie_year}, reflecting a surge in AI agent programming research following the widespread adoption of LLMs.
Based on our systematic analysis, we developed a hierarchical classification of AI agentic programming systems along behavioral characteristics and system architectures, as shown in Figure~\ref{fig:taxonomy-overview}. The following section examines each category in detail.

\section{Taxonomy of AI Agentic Programming}

AI agentic programming is an emerging paradigm that equips LLM-based systems with autonomy, enabling them to plan, execute, and refine programming tasks over multiple steps. To provide structure to the diverse and fast-evolving landscape of agentic programming systems, this section introduces a taxonomy based on key behavioral and architectural dimensions. We categorize existing systems and approaches along these axes to clarify the design space and inform future development.

\subsection{Agentic Behaviour Dimensions}
\label{subsec:4.1}
This subsection defines the primary behavioural traits that differentiate agentic systems, forming the basis for a comparative classification.

\subsubsection{Reactivity vs. proactivity}
Reactive agents respond directly to user prompts or feedback without independent task planning. For example, GitHub Copilot reacts by instantly suggesting a function body based on context after users type a function header like \texttt{def initial}. It does not include subsequent steps like writing tests or checking generated code.
Proactive agents initiate sub-tasks, form execution plans, and re-evaluate decisions, often working autonomously over extended periods. For example, a proactive LLM-based agent can decompose a task into subtasks and execute them automatically. Users providing high-level instructions, such as ``\emph{Add a user authentication module},'' will generate the login logic, update the database, integrate it with the UI, and write corresponding unit tests.

\subsubsection{Single-turn vs. multi-turn execution}
Single-turn agents perform actions in response to individual prompts, often without preserving context. For example, classical GitHub Copilot responds to each prompt independently, without remembering past interactions. In contrast, multi-turn agents, such as GitHub Copilot Agent or Claude Opus 4 with tooling capabilities, maintain state across interactions, enabling iterative refinement, exploration, and goal pursuit. For instance, GitHub Copilot Agent can hold a conversation across multiple steps, remember earlier function names, and build a complete module through back-and-forth iterations between agents \cite{github_copilot}.

\subsubsection{Tool-augmented vs. standalone agents}
Some agents are tightly integrated with external tools (e.g., compilers, debuggers, browsers, test frameworks), allowing them to perform code execution, validation, and correction. Others operate solely within the LLM’s reasoning capabilities, limiting their interactivity and adaptability.

\subsubsection{Static vs. adaptive agents}
Static agents (e.g., GitHub Copilot and Tabnine~\cite{tabnine}) follow predefined workflows or heuristics. Adaptive agents modify their strategies using feedback from tools, user input, or environmental signals. Some employ learning mechanisms to improve over time. For example, GitHub Copilot Agent adapts its approach when test failures occur, revising its implementation or replanning subtasks.

\subsection{Agent System Categories}
\label{subsec:4.2}
We now present a classification of current AI agentic programming systems, organised by their core functionality and architectural patterns.

\subsubsection{Interactive code assistants}
These are among the most widely adopted applications of LLMs in software development. These systems assist developers by providing code completions, inline documentation, editing suggestions, and simple refactorings. They are typically integrated directly into editors and IDEs, where developers interact with the underlying LLMs either through chat-like interfaces or by selecting code or comments using mouse-based interactions.

GitHub Copilot \cite{github_copilot} and Cursor \cite{cursor}  are two representative examples of LLM-based code assistants. GitHub Copilot, originally developed based on Codex trained on GitHub code repositories, offers context-aware code completions across multiple programming languages and is tightly integrated into popular IDEs like Visual Studio Code \cite{vscode}  and JetBrains. Cursor extends this functionality by embedding conversational interaction, maintaining memory of previous edits, and supporting structured command execution. Other notable systems include Amazon Q Developer \cite{amazon-q-developer}  and Tabnine \cite{tabnine}. Amazon Q Developer targets developers working in cloud ecosystems, offering language-specific completions, cloud API integration, and basic vulnerability detection. In contrast to Amazon Q Developer, Tabnine takes a privacy-first approach by deploying smaller local models trained on permissively licensed code, making it attractive for organizations and developers who do not want to send their code to a remote cloud. 

Implementations of these systems typically exhibit reactive behavior, responding to user input without initiating their own plans or taking proactive steps. Their interactions are generally single-turn, relying on the immediate context within the code editor rather than maintaining a persistent memory of past interactions or broader development goals. Most systems in this category are tightly coupled with development tools and offer real-time assistance that fits naturally within existing programming workflows. However, they are limited in autonomy, lacking the ability to decompose complex tasks, maintain long-term state, or coordinate multi-step development activities.

Despite these limitations, interactive code assistants serve as a foundational layer within the more recent agentic programming ecosystem. They are widely deployed, easy to integrate into everyday development practices, and offer immediate value to developers. Furthermore, some recent systems, such as Cursor and GitHub Copilot Agent, are beginning to incorporate features like session-level memory, persistent context, and structured task execution, gradually bridging the gap between reactive code assistants and more autonomous, multi-turn agents.

These distinctions are captured in our taxonomy (see Table~\ref{tab:taxonomy}), where interactive code assistants are compared with other categories of agentic systems across key dimensions, including autonomy, memory scope, tool integration, reasoning complexity, and interaction model.

\subsubsection{Autonomous task-oriented agents}
These agents perform multi-step programming tasks with minimal human intervention, often maintaining control over the entire development process from requirement interpretation to code generation and validation. They can plan, execute, and revise their own workflows in response to intermediate results or changing task requirements. Many integrate external tools such as debuggers, package managers, or search engines, enabling them to gather information, resolve errors, and optimize code without direct user guidance. 
Examples include GPT-5 \cite{openai2025gpt5introducing}, which functions as an end-to-end collaborator for long-horizon code generation and debugging; Claude Opus 4 \cite{anthropic_opus4}, engineered for sustained agentic workflows with strong long-context and tool-integration capabilities; Google Jules \cite{Jules}, which can sandbox repositories, propose diffs, and execute verified changes on real projects; DevGPT \cite{devgpt2024}, an open-source pipeline that converts tickets into actionable code and CI artifacts; Kimi K2 \cite{team2025kimi}, a model optimized for agentic coding proficiency; and Gemini 2 \cite{comanici2025gemini25pushingfrontier}, which emphasizes multimodal inputs and built-in planning modes.
These agents are often proactive in suggesting next steps, adaptive to new inputs, and capable of maintaining continuity across extended sessions through persistent memory or context management mechanisms.

\subsubsection{Planning-centric agents}
Planning-centric agents approach problem-solving as a two-phase process: first, structured task decomposition, where high-level goals are broken down into smaller, more manageable steps; and second, execution monitoring, in which results are evaluated and the plan is adjusted accordingly. This approach improves the handling of long-horizon tasks. 
For example, CAMEL \cite{li2023camel} employs two agents in a role-playing setup, a ``user'' agent and an ``assistant'' agent, which collaboratively refine goals and strategies, producing plans that downstream code generators can execute.
Voyager \cite{wang2023voyager} demonstrates this paradigm by continuously exploring an open-ended environment, generating executable skills, and refining them into reusable plans for long-term adaptability. 
Similarly, CodePlan \cite{wen2024codeplan} introduces code-form planning, where structured pseudocode serves as an explicit intermediate representation to decompose complex problems into executable steps.
These agents are typically multi-turn, memory-enabled, and trading speed for robustness.

\subsubsection{Multi-agent and collaborative systems}
Multi-agent and collaborative systems extend agent-based programming by introducing multiple specialized agents that coordinate to solve complex software engineering tasks. This approach draws inspiration from human software teams, where each member has a distinct role, such as requirements analysis, coding, testing, or documentation, and communication protocols are established to ensure progress toward shared objectives.
For example, SWE-Agent \cite{yang2024swe} employs multiple role-specific LLM agents: an ``Architect'' agent for high-level design, a ``Coder'' agent for implementation, and a ``Reviewer'' agent for quality assurance, which are connected through structured dialogue and shared memory. 
ChatDev \cite{qian2024chatdevcommunicativeagentssoftware} follows this paradigm by simulating an end-to-end software company, where agents take on roles such as CEO, CTO, and programmers to collaboratively design, implement, and test applications. 
Furthermore, AutoCodeRover \cite{zhang2024autocoderover} extends collaboration into real-world repositories, orchestrating specialized agents that autonomously navigate, edit, and validate source code across complex multi-file projects.

\subsection{Summary and Comparative Table}
\label{subsec:4.3}
We conclude this section with a comparative summary (Table~\ref{tab:taxonomy}) of representative systems across the behavioural dimensions and categories introduced above. This taxonomy provides a lens to understand the capabilities and limitations of existing approaches and can guide the design of future systems.

% Example placeholder for the table
\begin{table}[t]
\centering
\caption{Comparison of representative AI agentic programming systems.}
\label{tab:taxonomy}
\scriptsize
\begin{tabular}{llcccc}
\toprule
\textbf{System} & \textbf{Category} & \textbf{Proactivity} & \textbf{Multi-turn} & \textbf{Tool Use} & \textbf{Adaptivity} \\
\midrule
\rowcolor{myred!70} GitHub Copilot \cite{github_copilot}       & IDE Assistant      & Reactive  & \xmark   & \cmark     & \xmark      \\
\rowcolor{myred!70} {Amazon Q Developer} \cite{amazon-q-developer}       & IDE Assistant      &  Reactive &   \cmark  &  \cmark     &   \cmark     \\
\rowcolor{myred!70} Tabnine \cite{tabnine}              & IDE Assistant      & Reactive  & \xmark   & \xmark      & \xmark      \\
\rowcolor{myred!70} Cursor \cite{cursor}               & IDE Assistant      & Reactive  & \xmark   & \cmark     & \xmark      \\
\rowcolor{myorange} Claude Opus 4-powered agent \cite{anthropic_opus4}        & Task-oriented      & Proactive & \cmark  & \cmark     & \cmark     \\
\rowcolor{myorange} Google Jules \cite{Jules}         & Task-oriented      & Proactive & \cmark  & \cmark     & \cmark     \\
\rowcolor{myorange} DevGPT \cite{devgpt2024}              & Task-oriented      & Proactive & \cmark  & \cmark     & \cmark     \\
\rowcolor{myorange} KimI K2-powered agent \cite{team2025kimi}              & Task-oriented      & Proactive & \cmark  & \cmark     & \cmark     \\
\rowcolor{myorange} Gemini 2-powered agent \cite{comanici2025gemini25pushingfrontier}             & Task-oriented      & Proactive & \cmark  & \xmark      & \xmark \\
\rowcolor{myyellow} Voyager \cite{wang2023voyager}              & Planning Agent     & Proactive & \cmark  & \cmark     & \cmark     \\
\rowcolor{myyellow} CAMEL \cite{li2023camel}                & Planning Agent     & Proactive & \cmark  & \xmark      & \xmark      \\
\rowcolor{myyellow} CodePlan \cite{wen2024codeplan}             & Planning Agent     & Proactive & \cmark  & \xmark      & \xmark      \\
\rowcolor{mygreen} ChatDev \cite{qian2024chatdevcommunicativeagentssoftware}              & Multi-agent System & Proactive & \cmark  & \cmark     & \xmark      \\
\rowcolor{mygreen} AutoCodeRover \cite{zhang2024autocoderover}        & Multi-agent System & Proactive & \cmark  & \cmark     & \cmark     \\
\rowcolor{mygreen} SWE-Agent \cite{yang2024swe}            & Multi-agent System & Proactive & \cmark  & \cmark     & \cmark     \\
\bottomrule
\end{tabular}
\end{table}

\subsection{Cost and Token Consumption Model}
\label{subsec:cost-token-model}

While recent LLMs show impressive capabilities in software engineering tasks, their real-world applicability is often constrained by cost considerations. This cost is typically measured in terms of \emph{tokens consumed per US dollar} for both input and output, along with additional expenses incurred by extended reasoning strategies such as Chain-of-Thought (CoT) and tool-augmented workflows.

\subsubsection{Pricing dimensions.} 
Commercial providers generally price LLM usage by input and output tokens, with rates varying across model families. Some, such as OpenAI’s GPT-5, offer multiple service tiers (Standard, Mini, Nano, Pro) with different pricing, context lengths, and throughput limits. Table~\ref{tab:llm-pricing} summarizes representative pricing.

\begin{table}[t]
\centering
\caption{Example token pricing for LLMs used in coding tasks (USD per 1M tokens, as of Sep.\ 2025).}
\label{tab:llm-pricing}
\scriptsize
% \begin{adjustbox}{width=\textwidth,center}
\begin{tabular}{lrrr}
\toprule
\textbf{Model} & \textbf{Input(\$)} & \textbf{Output(\$)} & \textbf{Context Window} \\
\midrule
GPT-5 (Standard) & 1.25 & 10.00 & 256k \\
\rowcolor{gray!10} GPT-5 Mini & 0.25 & 2.00 & 128k \\
GPT-4 variants (e.g., 4o) & 2.50 & 10.00 & 128k \\
\rowcolor{gray!10} Claude 4 Opus & 15.00 & 75.00 & 200k \\
Gemini 2.5 Pro & 1.25 & 10 & 1M \\
\rowcolor{gray!10} Grok 4 & 3.00 & 15.00 &  256k \\
DeepSeek R1-0528 & 0.55 & 2.19 & 160k \\
\rowcolor{gray!10} Kimi K2 & 0.15 & 2.50 & 128k \\
Qwen3-235B-A22B & 0.22 & 0.88 & 128k \\
\rowcolor{gray!10} Qwen3-Coder-480B-A35B-Instruct & 4.5 & 13.5 & 256k \\
\rowcolor{gray!10} Solar-Pro & 0.30 & 0.30 & 128k \\
Openhands-LM-32B-v0.1 & 2.6 & 3.4 & 128k \\
\rowcolor{gray!10} Devstral-Small & 0.10 & 0.30 & 128k \\
Devstral-Medium & 0.40 & 2.00 & 128k \\
\bottomrule
\end{tabular}
\end{table}

\subsubsection{Impact of reasoning strategies.}
To make the cost-performance trade-offs more concrete, we draw on the ``Agentic workflow for implementing a REST task'' example described earlier in Section~\ref{sec:background}. In that workflow, the agent receives a high-level natural language specification for a REST API endpoint, interprets the requirements, generates code, and iteratively tests the implementation until it passes the provided unit tests.

We consider three reasoning strategies applied to this scenario:

\cparagraph{Short reasoning.} The model produces the implementation in a single turn with minimal intermediate reasoning. For the REST API task, this means directly generating the endpoint code and tests without explicit planning or validation steps. This approach minimizes token usage and latency but risks missing subtle requirements.

\cparagraph{Standard CoT.} The model uses a fixed-depth chain of thought to plan the implementation. In the REST API case, this involves reasoning about request handling, data validation, and error responses before generating the code. This strategy consumes significantly more tokens, roughly twice as many, compared to the short reasoning strategy, but yields a higher likelihood of producing a correct implementation on the first attempt.

\cparagraph{Tool-augmented iterative reasoning.} The agent integrates code compilation and test execution into the workflow. After producing an initial version, it runs the tests, inspects any failures, and revises the code in subsequent turns. For the REST API example, this may involve multiple cycles of fixing logic errors, adjusting request parsing, and refining edge-case handling. While this maximizes accuracy and robustness, it also increases token consumption and wall-clock time substantially due to repeated code generation and analysis.

In practice, the optimal choice depends on the cost-performance budget of the project. For time-sensitive or budget-constrained environments, a hybrid approach can offer a more effective balance.

\section{Challenges}

AI agentic programming introduces a promising and complex shift in how software is developed, relying on the autonomous capabilities of LLMs. Despite recent progress, several technical and conceptual challenges remain that hinder the deployment of robust, scalable, and trustworthy agentic systems~\cite{sapkota2025aiagentsvsagentic, acharya2025agentic}.

\subsection{Evaluation and Benchmarking}

A variety of benchmarks and open-source toolkits~\cite{liu2023agentbench, qin2023toolllmfacilitatinglargelanguage} have been proposed to evaluate the capabilities of LLM-based agents across different tasks. Despite this progress, existing benchmarks commonly used for assessing coding agents, such as HumanEval~\cite{zheng2023codegeex} and SWE-Bench~\cite{jimenez2023swe}, may still be inadequate for capturing the full complexity of real-world software engineering workflows.

Most of these benchmarks primarily focus on small, self-contained problems, often restricted to a small number of programming languages, such as Python~\cite{chen2021evaluatinglargelanguagemodels}, and generally lack support for interactive, multi-turn, or tool-integrated tasks~\cite{liu2023agentbench}. In contrast, practical agentic systems are expected to operate over large, modular codebases, interface with third-party libraries, manage build workflow pipelines, and respond dynamically to user feedback or runtime tool outputs.
As LLM-based systems increasingly incorporate reinforcement learning~\cite{ouyang2022training,wang2024survey} and more advanced planning mechanisms, future benchmarks should reflect this integration. Table~\ref{tab:swebench-tasks} shows the characteristics of tasks in SWE-Bench.
Although SWE-Bench introduces project-level repositories and leverages unit tests and continuous integration (CI) for evaluation, its scope is restricted to Python.
Most tasks are function- or module-level, with no support for multi-turn feedback, third-party library usage, or build pipeline management.
Even for project-level tasks, interaction is limited to binary pass/fail outcomes, providing only minimal support for realistic multi-step or tool-integrated software engineering workflows.
% \FIXME{FIXME!!!!!!!look into SWE-Bench as this includes project-level code bases. It is possible that SWE-Bench only requires solving problems within individual modules/functions. Provide a discussion on this, potentially with a table showing the complexity of the problems to be tested. }

Moreover, there is a noticeable absence of evaluation frameworks designed for emerging complex use cases, such as those involving interactions with compilers and debuggers~\cite{wang2024executablecodeactionselicit}, where agents must reason about low-level program behavior, perform iterative transformations, or track state across toolchains. The lack of such domain-specific benchmarks presents a significant gap in evaluating agent performance under realistic conditions.

\begin{table}[t]
\centering
\caption{SWE-Bench characteristics.}
\label{tab:swebench-tasks}
\scriptsize
\begin{tabular}{lrrrrrr}
\toprule
\textbf{Task type} & \textbf{Proportion} &\textbf{Languages} & \textbf{Interaction } & \textbf{Multi-turn feedback} & \textbf{Library integration} & \textbf{Build pipeline} \\
\midrule
Function-level& ~65\% & Python & Unit tests & \xmark & \xmark & \xmark \\
\rowcolor{gray!10} Module-level & ~25\% & Python & Tests and CI & \xmark & \xmark & \xmark \\
Project-level & <10\% & Python & Tests and CI & Limited (pass/fail) & \xmark &  \xmark\\

\bottomrule
\end{tabular}
\end{table}

\subsection{Communication Protocols for Multi-Agent Systems}
Early protocols enabled one-to-one agent–to–tool interactions but did not support direct agent-to-agent communication~\cite{MCP}. Current practice often relies on heterogeneous web service protocols and adapters, which provide interoperability but introduce high latency, bandwidth overhead, and limited scalability~\cite{naik2016web}. 
Recent work~\cite{jie_2012TSC, A2A, Eurosys25_verma} explores group session models, where each execution unit functions as a session or group agent coordinating multiple services, and standards efforts propose unified message formats and session semantics. However, the lack of a common protocol, inefficiencies of adapters, and challenges in managing dynamic multi-agent sessions remain open problems for scalable and dependable agentic systems.

\subsection{Domain-specific Models for Agents}

Generic coding agents often struggle in domain-specific environments such as embedded systems, high-performance computing, optimization, or formal software verification~\cite{lecong2025llmsreasonprogramsemantics, loughridge2024dafnybench}. These domains typically impose stricter operational constraints and require deep integration with specialized APIs, toolchains, and domain knowledge resources that are often underrepresented in general-purpose training corpora.
To address these limitations, recent research has proposed domain-adapted models and task-specific learning strategies to accelerate agent performance in specialized settings~\cite{song2025injecting}. For instance, some approaches have begun incorporating compiler knowledge or security-specific patterns into LLM training pipelines~\cite{lin2025llmssecuritycoursesecuring, LLMCompiler}, enabling agents to reason more effectively about low-level program behavior or vulnerability patterns.
In the future, developing robust and adaptable domain foundation models will be a promising direction for enabling agents to operate reliably in complex software environments, such as LLMs pretrained or fine-tuned on domain-specific data, tools, and semantics.

\subsection{Safety and Privacy}

As agentic systems gain increasing autonomy, so does the potential for unsafe behavior. Unlike traditional tools~\cite{Checkmarx}, agentic systems can invoke external tools, perform structural code modifications, and even commit changes without direct human oversight~\cite{triedman2025multiagentsystemsexecutearbitrary, zhang2025agentsafetybenchevaluatingsafetyllm}. These capabilities introduce significant risks, including the possibility of introducing subtle bugs, propagating unsafe patterns, or violating security constraints.
A critical future direction involves ensuring that agentic systems can protect users and data. For example, when agents visit private repositories or are deployed in cloud-integrated environments, future models may need built-in controls to restrict access to sensitive project data.
Further, malicious prompts, poisoned APIs, or compromised toolchains can mislead agents into executing unsafe behaviors~\cite{debenedetti2024agentdojodynamicenvironmentevaluate, zhan2024injecagentbenchmarkingindirectprompt}. Future research should prioritize the design of secure protocols for agent collaboration, including authentication between agents, validation of tool outputs, and detection of anomalous actions. 
Also, agents must be capable of explaining their reasoning, flagging uncertainties, and allowing developers to understand and revise with minimal effort. Building safety and privacy into the foundation of agentic architectures is essential.

\subsection{Toolchain Integration and Programming Language Design}

One fundamental challenge is the incompatibility between existing software tools and the needs of autonomous coding agents. Most programming languages, compilers, debuggers, and development environments were designed for human developers~\cite{rorseth2025ladybug}. They emphasize usability and readability over structured, machine-consumable feedback. As a result, agents often struggle to diagnose failures, trace the consequences of code transformations, or interpret build errors~\cite{kim2024llm}.
For example, compilers typically report transformation failures or semantic conflicts in the form of coarse error messages, providing little insight into why an optimization was blocked or how a type error emerged~\cite{li2025review}. Languages similarly prioritize human readability over machine-negotiated meaning, while compilers conceal internal reasoning to avoid overwhelming human users. These design choices, while historically effective, now limit the ability of AI coding agents to construct safe, efficient, and adaptive software.

To enable agentic workflows, toolchains should evolve to expose richer intermediate representations, transformation traces, and structured feedback interfaces. Equally important, programming languages should incorporate annotations and agent-aware interfaces that make developer intent explicit, allowing automated reasoning to be guided by semantic contracts rather than inferred heuristics. Together, these advances would transform compilers and languages from opaque tools into collaborative infrastructures capable of supporting autonomous agents at scale.

\subsection{Scalable Memory}

Agentic programming systems must maintain coherence and reasoning over long-running tasks involving multiple iterations, tools, and contextual dependencies. However, current LLMs are limited by fixed context windows and lack persistent, structured memory mechanisms~\cite{wang2023augmenting}.
Realistic software tasks may require agents to store and reason over evolving states, feedback logs, intermediate plans, and prior actions~\cite{qian2025memorag}. Without hierarchical and queryable memory systems, agents risk repeating errors, forgetting past successes, or producing inconsistent results.
Emerging solutions such as retrieval-augmented generation and memory summarization offer partial relief, but they remain inadequate for complex, multi-session workflows~\cite{maharana2024evaluating}. Future research can explore memory architectures that differentiate short-term interactions, mid-term subgoals, and long-term domain knowledge.

\section{Opportunities and Future Directions}

\begin{figure}[t]
    \centering

  \begin{adjustbox}{width=\linewidth}
    % 仅 tikzpicture；由 main.tex 负责加载 tikz/xcolor/颜色定义
\begin{tikzpicture}[
  >={Stealth[length=5pt,width=6pt]},
  box/.style={
    draw=black!70, rounded corners=3pt,
    align=center, anchor=west,          % 用 west 锚点，保证同列左边缘对齐
    inner sep=6pt, minimum height=11mm
  },
  arr/.style={->, line width=0.9pt, draw=black}
]

% ===== 列宽（可按需修改） =====
\newcommand{\wA}{28mm} % 第1列
\newcommand{\wB}{36mm} % 第2列
\newcommand{\wC}{36mm} % 第3列
\newcommand{\wD}{36mm} % 第4列
\newcommand{\wE}{38mm} % 第5列

% 间距
\newcommand{\colsep}{10mm}  % 列间距
\newcommand{\rowsep}{18mm}  % 行间距

% ===== 定义每一列的 X 坐标（从 col1 开始依次累加宽度+间距） =====
\coordinate (col1) at (0,0);
\coordinate (col2) at ($(col1)+(\wA+\colsep,0)$);
\coordinate (col3) at ($(col2)+(\wB+\colsep,0)$);
\coordinate (col4) at ($(col3)+(\wC+\colsep,0)$);
\coordinate (col5) at ($(col4)+(\wD+\colsep,0)$);

% ===== 定义每一行的 Y 坐标（向下递减） =====
\coordinate (row1) at (0,0);
\coordinate (row2) at ($(row1)-(0,\rowsep)$);
\coordinate (row3) at ($(row2)-(0,\rowsep)$);
\coordinate (row4) at ($(row3)-(0,\rowsep)$);
\coordinate (row5) at ($(row4)-(0,\rowsep)$);
\coordinate (row6) at ($(row5)-(0,\rowsep)$);
\coordinate (row7) at ($(row6)-(0,\rowsep)$);
\coordinate (row8) at ($(row7)-(0,\rowsep)$);

% ====== 第 1 行 ======
\node[box, fill=beaublue!60,  text width=\wA] (r1c1) at (col1 |- row1) {Software Compatibility\\(Section \ref{subsec:future-Integrating})};
\node[box, fill=myorange,     text width=\wB] (r1c2) at (col2 |- row1) {Structured feedback on failures};
\node[box, fill=myorange,     text width=\wC] (r1c3) at (col3 |- row1) {Direct IR interaction};
\node[box, fill=myorange,     text width=\wD] (r1c4) at (col4 |- row1) {Standardized compiler interfaces};
\node[box, fill=myyellow!100, text width=\wE] (r1c5) at (col5 |- row1) {How do coding agents integrate with tools?};

% ====== 第 2 行 ======
\node[box, fill=beaublue!60,  text width=\wA] (r2c1) at (col1 |- row2) {Context Management\\(Section \ref{subsec:future-Scalable})};
\node[box, fill=myorange,     text width=\wB] (r2c2) at (col2 |- row2) {Structured storage and recall of evolving task information};
\node[box, fill=myorange,     text width=\wC] (r2c3) at (col3 |- row2) {Conditioned retrieval from task state and tool feedback};
\node[box, fill=myorange,     text width=\wD] (r2c4) at (col4 |- row2) {Program state tracing and replay};
\node[box, fill=myyellow!100, text width=\wE] (r2c5) at (col5 |- row2) {How do coding agents manage memory and contextual information?};

% ====== 第 3 行 ======
\node[box, fill=beaublue!60,  text width=\wA] (r3c1) at (col1 |- row3) {Benchmarks\\(Section \ref{subsec:future-Evaluation})};
\node[box, fill=myorange,     text width=\wB] (r3c2) at (col2 |- row3) {Heavily biased training corpus};
\node[box, fill=myorange,     text width=\wC] (r3c3) at (col3 |- row3) {Large absence of real-world tasks};
\node[box, fill=myorange,     text width=\wD] (r3c4) at (col4 |- row3) {Lack of reflecting interactive and iterative programming workflow};
\node[box, fill=myyellow!100, text width=\wE] (r3c5) at (col5 |- row3) {How to comprehensively evaluate coding agents?};

% ====== 第 4 行 ======
\node[box, fill=beaublue!60,  text width=\wA] (r4c1) at (col1 |- row4) {Human-AI Collaboration\\(Section \ref{subsec:future-Human})};
\node[box, fill=myorange,     text width=\wB] (r4c2) at (col2 |- row4) {Frameworks for responsibility allocation};
\node[box, fill=myorange,     text width=\wC] (r4c3) at (col3 |- row4) {Handle uncertainty; Ecosystem integration};
\node[box, fill=myorange,     text width=\wD] (r4c4) at (col4 |- row4) {Coordination in multi-user environments};
\node[box, fill=myyellow!100, text width=\wE] (r4c5) at (col5 |- row4) {How do coding agents collaborate with human?};

% ====== 第 5 行 ======
\node[box, fill=beaublue!60,  text width=\wA] (r5c1) at (col1 |- row5) {Domain-specific Agents\\(Section \ref{subsec:future-Domain})};
\node[box, fill=myorange,     text width=\wB] (r5c2) at (col2 |- row5) {Adaptive prompting, fine-tuning, and behavior modulation};
\node[box, fill=myorange,     text width=\wC] (r5c3) at (col3 |- row5) {Reasoning-loop integration of domain tools and diagnostics};
\node[box, fill=myorange,     text width=\wD] (r5c4) at (col4 |- row5) {Runtime domain detection and adaptation};
\node[box, fill=myyellow!100, text width=\wE] (r5c5) at (col5 |- row5) {How to develop domain-specific coding agents?};

% ====== 第 6 行 ======
\node[box, fill=beaublue!60,  text width=\wA] (r6c1) at (col1 |- row6) {Safety, Alignment and Trust\\(Section \ref{subsec:future-Safety})};
\node[box, fill=myorange,     text width=\wB] (r6c2) at (col2 |- row6) {Agent behavior grounding};
\node[box, fill=myorange,     text width=\wC] (r6c3) at (col3 |- row6) {Risk detection and reduction};
\node[box, fill=myorange,     text width=\wD] (r6c4) at (col4 |- row6) {Explainable and self-reflective agents};
\node[box, fill=myyellow!100, text width=\wE] (r6c5) at (col5 |- row6) {How to ensure safer, ethical, and trustworthy coding agents?};

% ====== 第 7 行 ======
\node[box, fill=beaublue!60,  text width=\wA] (r7c1) at (col1 |- row7) {Multi-Agent Collaboration\\(Section \ref{subsec:future-Multi})};
\node[box, fill=myorange,     text width=\wB] (r7c2) at (col2 |- row7) {Native protocol support for multi-agent systems};
\node[box, fill=myorange,     text width=\wC] (r7c3) at (col3 |- row7) {Session management for dynamic coordination};
\node[box, fill=myorange,     text width=\wD] (r7c4) at (col4 |- row7) {Semantics-aware routing, conflict resolution, and consistency};
\node[box, fill=myyellow!100, text width=\wE] (r7c5) at (col5 |- row7) {How to design standardized protocols for multi-agent collaboration?};

% ====== 第 8 行 ======
\node[box, fill=beaublue!60,  text width=\wA] (r8c1) at (col1 |- row8) {System Support\\(Section \ref{subsec:future-System})};
\node[box, fill=myorange,     text width=\wB] (r8c2) at (col2 |- row8) {System resource scheduling (CPU, GPU, memory)};
\node[box, fill=myorange,     text width=\wC] (r8c3) at (col3 |- row8) {Reducing overhead in communication with environments};
\node[box, fill=myorange,     text width=\wD] (r8c4) at (col4 |- row8) {Coordinating tasks and resolving conflicts in collaborative codebases};
\node[box, fill=myyellow!100, text width=\wE] (r8c5) at (col5 |- row8) {How to enhance system-level support for coding agents?};

% ===== 水平箭头（同一行天然同 y） =====
\foreach \r in {1,2,3,4,5,6,7,8}{
  \foreach \c/\n in {1/2,2/3,3/4,4/5}{
    \draw[arr] (r\r c\c.east) -- (r\r c\n.west);
  }
}

\end{tikzpicture}
  \end{adjustbox}
    
    \caption{Summary of potential future research directions of AI agentic programming.}
    \label{fig:opportunities}
\end{figure}
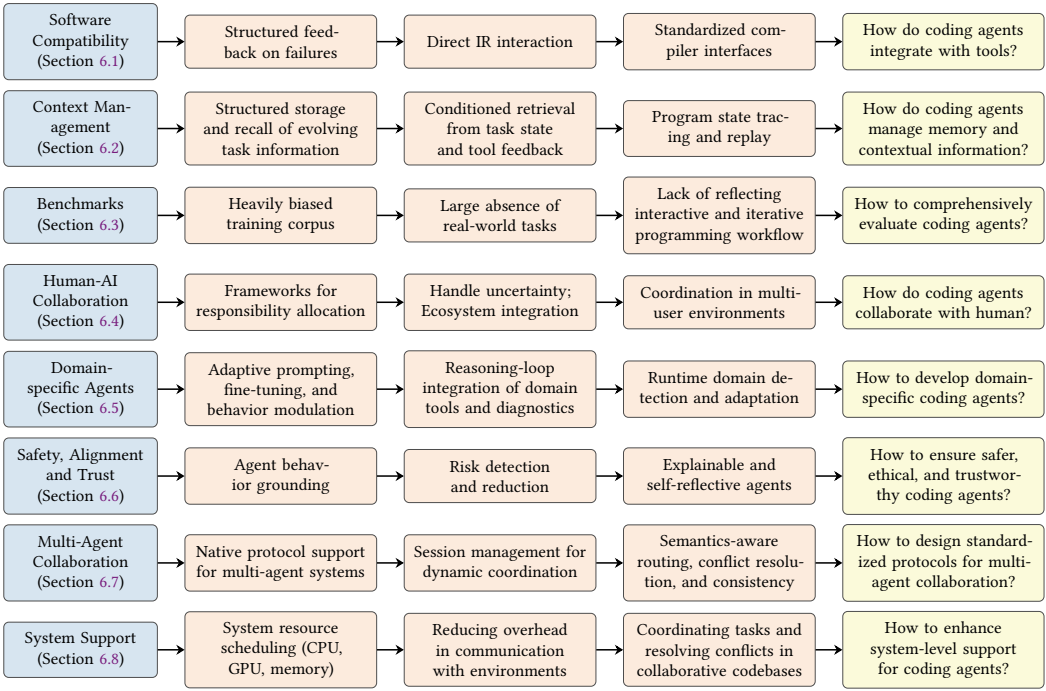

AI agentic programming represents a fast-evolving research frontier that intersects artificial intelligence, programming languages, and software engineering. While recent advances have demonstrated promising capabilities, significant challenges remain in realizing robust, efficient, and trustworthy agentic systems. In this section, we outline several key opportunities and open research directions that can shape the future of this field. An overview is provided in Figure~\ref{fig:opportunities}.

\subsection{Integrating Coding Agents with Tools\label{subsec:future-Integrating}}

Existing AI coding agents typically orchestrate LLMs with loosely integrated toolchains and basic memory mechanisms. These ad hoc designs often lack robustness, scalability, and generalization across programming tasks. Advancing agent architectures will require moving beyond simple prompt-response patterns toward more modular, structured systems that support reasoning, tool interaction, planning, and verification.

A promising direction is to rethink how programming languages, compilers, and testing frameworks, which are traditionally built for human developers, can be redesigned to support AI coding agents. For example, instead of emitting opaque diagnostics, compilers could provide structured feedback explaining why certain optimizations (e.g., vectorization or inlining) fail \cite{maleki2011evaluation,ashouri2018survey,siso2019evaluating}. These could include semantic barriers like unresolved aliasing, ambiguous data/control flow, or missing annotations~\cite{bi2024iterative}, enabling agents to revise code more precisely.
Beyond diagnostics, compilers can help agents track state across iterations. Feedback on which edits introduced errors, failed assertions, or performance regressions would enable agents to reason over the change history and adjust strategies accordingly.

Opening compiler internals also presents a valuable opportunity. Coding agents could interact directly with intermediate representations (IRs), such as LLVM IR~\cite{lattner2004llvm} or MLIR~\cite{lattner2021mlir}, to reason about program structure, verify transformations, or perform static analysis at a semantic level. Compiler APIs and language servers (e.g., Clang’s LibTooling, the Language Server Protocol) already expose ASTs, symbol tables, and refactoring tools, but a wider adoption may require standardized, introspective interfaces across compilers.

At the programming language level, agent-aware extensions or annotations could further improve interaction. Developers might use domain-specific languages, embedded contracts, or even natural language comments, e.g., “sort the elements of input array $x$”, to convey intent. This could guide synthesis, verification, or debugging. Likewise, compilers might expose symbolic summaries of control flow, memory access patterns, or performance profiles to inform multi-step agent reasoning.

Tighter integration with runtime systems also offers opportunities. For instance, agents can dynamically insert instrumentation or launch profiling runs, then use the results to inform optimization choices. Coupling these capabilities with autotuning frameworks~\cite{chen2018tvm,ansel2014opentuner} would expand the design space while preserving correctness and safety.
Finally, advances in structured code representations, such as ASTs, graph-based IRs, and semantic embeddings, offer a foundation for more powerful agent reasoning. Combining LLMs with graph neural networks or neuro-symbolic systems could improve generalization and support cross-language, cross-target understanding.

\subsection{Scalable Memory and Context Management\label{subsec:future-Scalable}}

A key capability of agentic programming lies in managing memory and contextual information across tasks that involve long context reasoning and multiple iterations. Unlike traditional code generation, which typically follows a single pass prompt to solution model, agentic workflows for solving real-world software engineering problems involve multiple steps, iterative refinement, and integration with external tools and development environments~\cite{cursor,wang2024openhands,openai2021codex,github_copilot}.

Consider an agent tasked with adding a new feature to an open source project, such as implementing a command-line flag to enable verbose logging. The agent must first analyze the existing codebase to locate the argument parsing logic, generate the required code changes, and update the logging behavior. If the updated code fails with a runtime error due to an uninitialized flag, the agent needs to debug the issue by inspecting stack traces, revise the code accordingly, and rerun the tests. Once the implementation is verified, the agent writes a commit message, creates a pull request with a summary of the changes, and links it to the relevant issue.

Throughout this process, the agent must persist and reason over a large and evolving context: the initial task description, previously generated code, compiler and runtime feedback, and version control metadata. Without the ability to store and recall this information in a structured way, the agent may repeat past mistakes, forget earlier successful changes, or submit incomplete solutions. Agentic programming, therefore, needs mechanisms for memory and context tracking that go beyond simple token limits, enabling coding agents to maintain continuity across extended interactions and tool usage.

As the memory footprint of LLMs grows linearly with input token length \cite{DBLP:conf/nips/VaswaniSPUJGKP17,kaplan2020scaling}, current LLM-based agents remain constrained by their context windows and lack persistent memory across a long sequence of iterations \cite{schick2023toolformer,liu2023lost,packer2023memgpt}. However, reasoning about real-world programs often requires modeling complex data structures and code context (like function calls) spanning across multiple files, which often exceeds typical context limits. Although some industry-scale LLMs claim to support million-token contexts~\cite{ding2023longnet,team2024gemini,ding2024longrope}, they often rely on random sampling techniques~\cite{hosseini2024efficient} and fail to leverage program structure or semantics effectively.

Therefore, an interesting direction is to design attention mechanisms that are guided by code structure, such as syntax trees, control flow graphs, or data dependencies. These structures can help agents focus more accurately on the most relevant parts of a program. While approaches like retrieval-augmented generation (RAG) \cite{lewis2020retrieval}, KV cache offloading \cite{lee2024infinigen,sun2024shadowkv}, and compression \cite{liu2024kivi,chang2025palu} provide partial solutions, they struggle to provide precise control over long-term dependencies, structured knowledge, and execution histories.

AI coding agents can also benefit from hierarchical memory models that distinguish between short-term interaction history, mid-term planning objectives (such as subgoals and intermediate decisions), and long-term knowledge. This long-term layer may include patterns of success or failure, reusable code templates, and observed tool behaviors. Such hierarchies can be dynamically updated and selectively queried using retrieval controllers or attention-based mechanisms. Additionally, memory summarization techniques could be explored to condense lengthy interaction histories into structured, semantically meaningful representations. For example, an agent might summarize a multi-turn session as a sequence of planning decisions and outcomes, highlighting key insights and interventions.

Another important area is the development of context-aware retrieval strategies that move beyond static similarity-based methods \cite{su2024dragin}. During debugging, for instance, an agent could retrieve not only the most recent error message but also similar past failures, proposed fixes, relevant test cases, and their outcomes. Retrieval conditioned on task state and tool feedback would significantly improve the agent's ability to reason under uncertainty.

Structured mechanisms for program state tracing and replay may also enhance agent performance. By recording partial program states, tool outputs, and execution steps, agents can support backtracking, recovery from failure, and richer explanations. For example, an agent could explain how a specific code edit introduced a type error or why a particular memory access blocked loop vectorization. These capabilities are crucial for supporting causal reasoning and improving transparency.
Likewise, persistent memory across multiple code generation and refinement sessions will be essential for enabling agents to accumulate and refine knowledge over time. This may include long-term storage of project-specific context, interaction histories, usage patterns of tools, and models of user intent. Such memory infrastructure will allow for continual learning and increasing personalization of agent behavior.

In summary, effective memory and context management are foundational for scaling agentic programming systems. These capabilities are vital for advancing from reactive, prompt-driven assistants to autonomous, context-aware collaborators capable of sustained reasoning, adaptation, and long-term learning.

\subsection{Evaluation and Benchmarking\label{subsec:future-Evaluation}}

\begin{table}[t]
\centering
\scriptsize
\caption{\centering Benchmarks for evaluating LLMs and agentic systems on programming tasks.\\ Abbreviations: CP = Competitive Programming, BF = Bug Fixing, FC = Function Completion, CR = CLI Reasoning, PO = Performance Optimization.}
\vspace{-2mm}
\label{tab:benchmarks}
\begin{adjustbox}{center}
\begin{tabular}{llllll}
\toprule
\textbf{Benchmark} & \textbf{Source} & \textbf{Language} & \textbf{Task} & \textbf{Difficulty} & \textbf{Year} \\
\midrule
HumanEval      & Hand-written      & Python                 & FC         & Beginner      & 2021 \\
\rowcolor{gray!10} MBPP           & Crowd-sourced     & Python                 & FC         & Beginner      & 2021 \\
CodeContests   & CP                & Python/C++/Java        & FC         & Diverse       & 2022 \\
\rowcolor{gray!10} HumanEval-X    & Hand-written      & Python/C++/Java/JS/Go  & FC         & Intermediate  & 2023 \\
SWE-Bench      & GitHub Issues     & Python                 & BF         & Expert        & 2024 \\
\rowcolor{gray!10} SWE-bench M    & GitHub Issues     & JS                     & BF         & Diverse       & 2024 \\
LiveCodeBench  & CP (live)         & Python                 & FC, BF     & Diverse       & 2024 \\
\rowcolor{gray!10} Terminal-Bench  & Community-curated & Shell                  & CR         & Diverse       & 2025 \\
Spider 2.0     & Enterprise DB Apps& SQL                    & FC         & Expert        & 2025 \\
\rowcolor{gray!10} EffiBench-X    & Synthetic         & Python/C++/JS          & FC, BF     & Diverse       & 2025 \\
Web-Bench      & Web App Projects  & JS/TS/HTML/CSS/Python  & BF, FC     & Expert        & 2025 \\
\rowcolor{gray!10} ProjectEval    & Open-source Repos & Python/Java/C++/JS     & FC, BF, CR & Expert        & 2025 \\
TRAIL          & CP                & Python/Java/C++/JS     & BF, CR     & Expert        & 2025 \\
GSO          &  Open-source Repos          & Python/C/C++/Cython/Rust    &   PO   & Expert       & 2025 \\
\bottomrule
\end{tabular}
\end{adjustbox}

\footnotesize\textit{Note: "Diverse" difficulty indicates that the benchmark covers a wide range from beginner to expert-level tasks.}
\vspace{-2mm}
\end{table}

\begin{figure*}[t]
           \subfigure[GSO]{
    \begin{minipage}[t]{0.23\textwidth}
    \centering
    \includegraphics[width=\columnwidth]{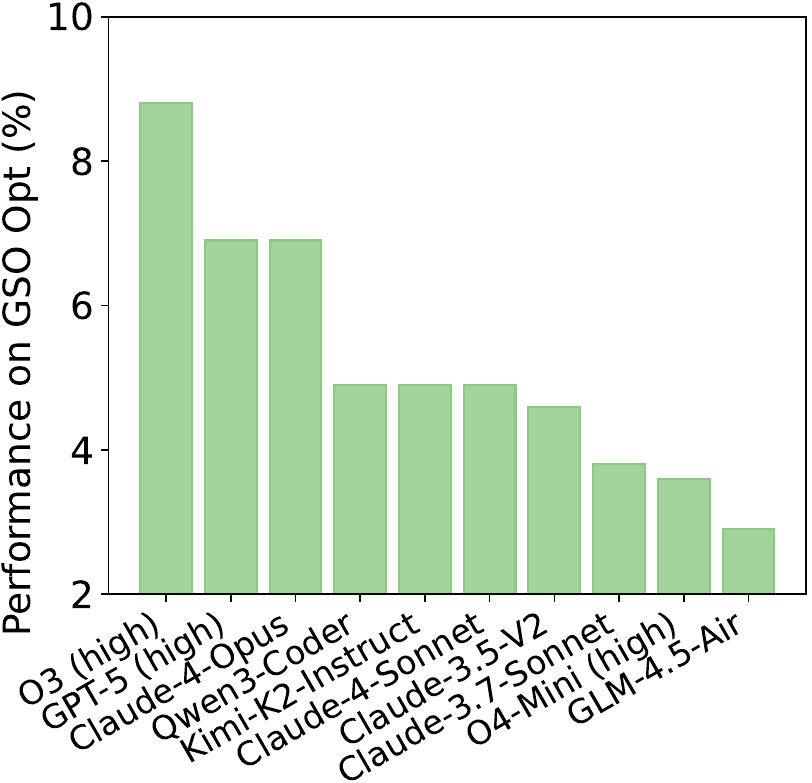}
    %\caption{fig1}
    \end{minipage}
    }
    \subfigure[LiveCodeBench]{
    \begin{minipage}[t]{0.23\textwidth}
    \centering
    \includegraphics[width=\columnwidth]{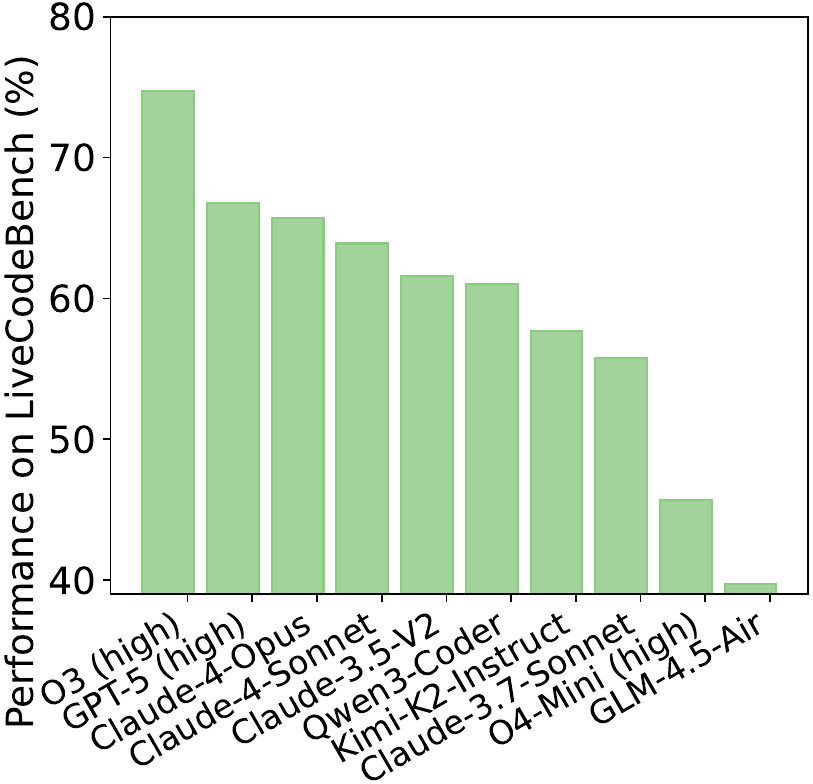}
    %\caption{fig2}
    \end{minipage}
    }
    \subfigure[Long Context Reasoning]{
    \begin{minipage}[t]{0.23\textwidth}
    \centering
    \includegraphics[width=\columnwidth]{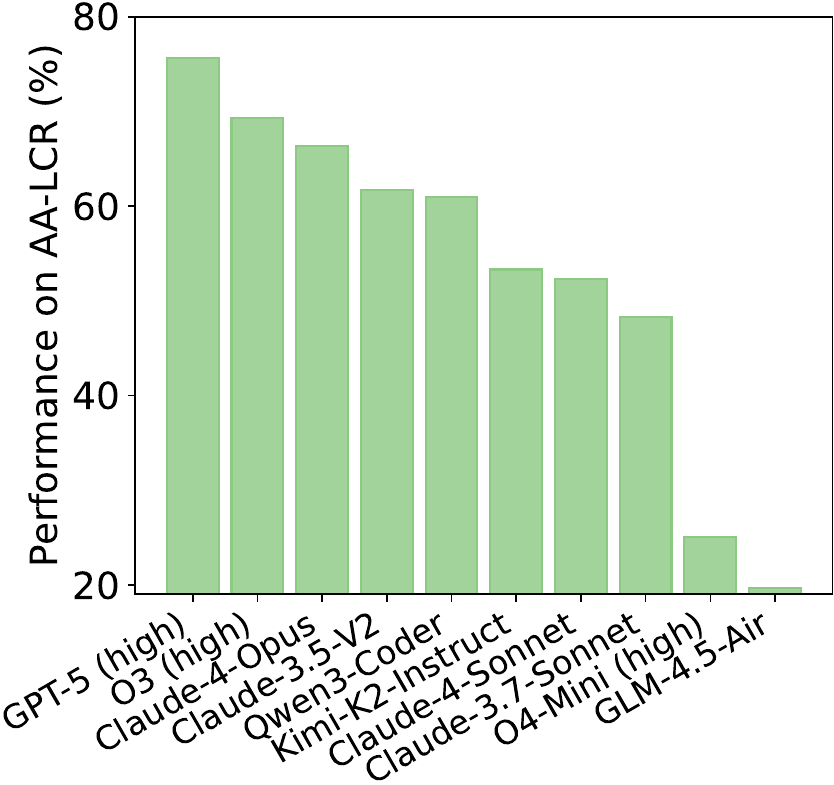}
    %\caption{fig3}
    \end{minipage}
    }
        \subfigure[Terminal-Bench]{
    \begin{minipage}[t]{0.23\textwidth}
    \centering
    \includegraphics[width=\columnwidth]{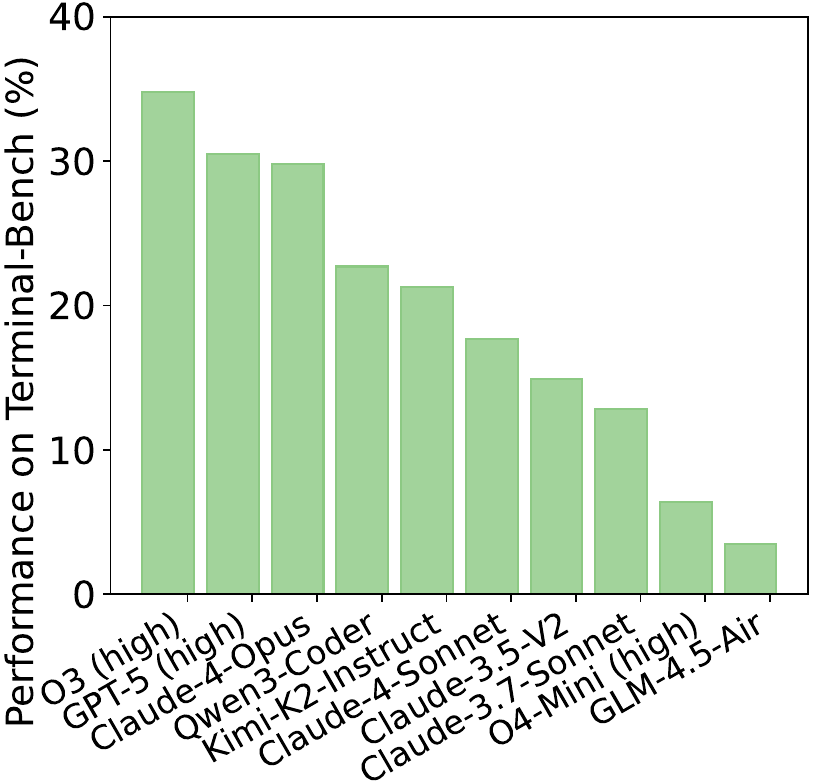}
    %\caption{fig4}
    \end{minipage}
    }
    \centering
\vspace{-2mm}
    \caption{Performance comparison of the top 10 language models across four representative benchmarks, including software optimization (GSO), code generation (LiveCodeBench), long-context reasoning (AA-LCR), and system-level tasks (Terminal-Bench).} \label{fig:res}

\end{figure*}
% \begin{table}[t]
% \centering
% \scriptsize
% \caption{AI Intelligence Score (across mathematics, science, coding, and reasoning) and Speed (Output tokens per second) for LLMs on LivecodeBench}
% \label{tab:models-res}
% \begin{tabularx}{\textwidth}{l *{14}{>{\centering\arraybackslash}X}}
% \toprule
% \textbf{Metric} 
% & GPT-5 & GPT-4o & Claude 4 & Gemini 2.5 Pro & Grok 4 & DeepSeek R1 & Kimi K2 & Qwen3 & Solar-Pro & Devstral  & Llama 4  & Magistral \\
% \midrule
% AI score 
% & 69 & 65 & 59 & 65 & 68 & 59 & 49 & 64 & 43 & 31 &  42 & 36 \\
% Speed 
% & 102 & 148 & 100 & 153 & 76 & 21 & 45.2 & 24 & 54 & 44  & 174 & 51 \\
% \bottomrule
% \end{tabularx}
% \end{table}

% More benchmarks:
% CodeContests+: High-Quality Test Case Generation for Competitive Programming
% HumanEval+: Is Your Code Generated by ChatGPT Really Correct? Rigorous Evaluation of Large Language Models for Code Generation
% MultiPL-E: A Scalable and Extensible Approach to Benchmarking Neural Code Generation

Table~\ref{tab:benchmarks} summarizes several widely used coding benchmarks for evaluating LLMs and agentic systems on programming tasks. These benchmarks vary in their origin, programming language coverage, task types, and difficulty levels. Commonly used datasets include \textit{HumanEval} \cite{openai2021codex}, \textit{HumanEval-X} \cite{zheng2023codegeex}, \textit{MBPP} \cite{austin2021program}, \textit{SWE-Bench} \cite{jimenez2023swe}, \textit{SWE-bench Multimodal} \cite{yang2024swe}, \textit{Terminal-Bench} \cite{tbench_2025}, \textit{LiveCodeBench} \cite{jain2024livecodebench}, \textit{CodeContests} \cite{DBLP:journals/corr/abs-2203-07814}, \textit{Spider2.0} \cite{lei2024spider}, \textit{EffiBench-X} \cite{qing2025}, \textit{Web-Bench} \cite{xu2025webbench}, \textit{ProjectEval} \cite{liu-etal-2025-projecteval}, \textit{TRAIL} \cite{deshpande2025trailtracereasoningagentic}, and \textit{GSO} \cite{shetty2025gso}.

% \FIXME{Populate Table~\ref{tab:code-benchmarks} with key benchmarks, including columns for source (e.g., GitHub issues, competitive programming), supported languages, task type (e.g., bug fixing, function completion, CLI reasoning), and difficulty level (e.g., beginner, intermediate, expert).}
% \FIXME{Consider including some results from https://artificialanalysis.ai/evaluations/livecodebench}

While these benchmarks have provided valuable insights into the capabilities of LLMs and agentic systems for code generation and bug fixing, they also have important limitations. For example, they are heavily biased toward a small set of programming languages - especially Python, which dominates the training corpus of code for current models \cite{twist2025llms,chai2024mceval}. This limits the generalizability of evaluation results to domains involving statically typed or domain-specific languages, such as C++ or Rust \cite{twist2025llms}. For example, Figure~\ref{fig:res} presents the performance of the top-10 LLMs across four representative benchmarks. The results reveal clear task-dependent differences. For example, all models still perform poorly on software optimization tasks, while they perform well on code generation tasks. Overall, performance gaps remain across all benchmarks, indicating that current LLMs still face significant challenges in reasoning, coding, and generalization to complex tasks.

Furthermore, many existing benchmarks focus on small, self-contained problems that may not be representative of real-world software engineering tasks. Realistic development scenarios often involve working with large, modular codebases, extensive use of third-party libraries, non-trivial build processes, and long-range dependencies across files and components. These characteristics are largely absent from current benchmark datasets, making it difficult to assess an agent's ability to scale or generalize. Another key limitation is that most benchmarks do not capture the interactive, iterative nature of agentic programming. In real-world settings, coding agents will need to collaborate with human developers \cite{wang2024openhands,cursor,openai2021codex,github_copilot}, receive intermediate feedback and confirmation, and rely on external tools such as compilers, debuggers, and test frameworks. Benchmarks that assume single-shot or non-interactive task completion fail to reflect the complexity of such multi-step, tool-augmented workflows.

Addressing these gaps will require developing more comprehensive and extensible evaluation frameworks. Future benchmarks should incorporate realistic tasks that reflect end-to-end development workflows, support multiple programming languages, and enable interaction with tools and human feedback loops. Metrics should go beyond functional correctness to include robustness, tool usage efficiency, recovery from failure, and the ability to incorporate feedback. Simulation environments and evaluation harnesses will also be important for reproducibility and fair comparison. 

\subsection{Human-AI Collaboration\label{subsec:future-Human}}
While a long-term vision of agentic programming is to automate the entire software development lifecycle, including writing, debugging, and testing code, near-term opportunities include extending the capabilities of current LLM-based coding assistants \cite{wang2024openhands,cursor,openai2021codex,github_copilot}. Rather than replacing human developers, these systems can act as collaborative partners, supporting workflows in which humans retain strategic oversight. Similar to pair programming \cite{hannay2009effectiveness}, LLM-based agents can assist by proposing ideas, detecting errors, suggesting improvements, and automating routine tasks, thereby augmenting human productivity.

Designing effective models for human-AI collaboration is a key research challenge. This involves creating user interfaces, interaction protocols, and responsibility-sharing frameworks suited for professional and team-based environments where coordination, trust, and efficiency are critical. Unlike traditional code generation tools, agentic systems operate in iterative, feedback-driven loops and make autonomous decisions based on intermediate results. This shifts the developer’s role from command giver to collaborator, opening new opportunities and challenges for co-creating software. To support this shift, agents must be transparent in their reasoning, responsive to human input, and capable of explaining their decisions, which calls for advances in interactive prompting~\cite{schulhoff2024prompt}, natural language explanations~\cite{creswell2022faithful}, and context-aware dialogue protocols~\cite{shah2025prompt,anthropic2024mcp}.

An interesting direction would be mixed initiative workflows, where control moves fluidly between human and agent. Developers can specify goals, constraints, or coding conventions, while agents generate scaffolding code, explore alternatives, or automate repetitive tasks. The developer then inspects, accepts, rejects, or refines the output. For example, an agent may suggest several refactoring options and generate test cases to validate them, leaving the final choice to the human. Such workflows must also support interruption and resumption, enabling agents to incorporate new input without losing context. This requires toolchains that track shared state across iterations, including goals, partial programs, and transformation history.

Finally, domain-aware collaboration can strengthen interpretability and safety. Agents should handle ambiguous or incomplete requirements by asking clarifying questions~\cite{zhang2024modeling,mu2024clarifygpt}, drawing on design documents, issue trackers, and domain knowledge to refine their reasoning. They should also provide clear explanations, cite relevant evidence, and highlight trade-offs to build trust, which is especially critical in domains such as finance, healthcare, and aerospace.

\subsection{Domain Specialization and Adaptability\label{subsec:future-Domain}}
While LLMs trained on extensive code corpora can demonstrate general programming capabilities~\cite{joel2024survey,zheng2024well}, they may struggle in specialized domains, such as generating Verilog code for hardware synthesis, which involves strict performance constraints, low-level abstractions, or domain-specific libraries and tooling.

Future research can improve agentic programming by exploring domain adaptive prompting, fine-tuning, and behavior modulation in areas such as embedded systems, data science, scientific computing, high-performance computing, and formal methods. In embedded systems programming~\cite{englhardt2024exploring}, for example, agents must reason about memory layout, timing constraints, and hardware-specific APIs, which are often underrepresented in general training data. Training on domain-specific code or adjusting planning strategies to prioritise correctness and safety over brevity may enhance performance. Integrating domain-specific tools and diagnostics into agents' reasoning, such as simulation engines~\cite{openfoam,permann2020moose,simulink}, static analyzers~\cite{codeql2020}, fuzzers~\cite{manes2019art}, or formal verification tools~\cite{woodcock2009formal}, can further improve robustness. Specialisation also supports interpretability and safety, enabling stricter validation, better use of formal specifications, and more reliable feedback in safety-critical domains such as aerospace, automotive, and healthcare.

Additionally, research into adaptive agent behavior can enable models to detect the domain of a task at runtime and adjust their prompting strategies, tool usage, and explanation styles accordingly. For example, an agent working on a financial modeling script might switch to using Pandas and SQL queries, while one dealing with real-time control code may emphasize low-latency function design and interrupt safety. This line of research will also open opportunities for personalisation at the developer or team level. Agents can learn preferences, coding conventions, and domain-specific heuristics from local repositories or past interactions, enabling more consistent and context-aware assistance.

\subsection{Safety, Alignment, and Trust\label{subsec:future-Safety}}

As agentic coding systems start taking on greater responsibility in software development, it becomes increasingly important to ensure that their behavior aligns with user intent, produces correct results, and avoids unintended changes~\cite{ouyang2022training,liu2024exploring}. Unlike traditional LLM-based code assistants, agentic systems can take multi-step actions, use external tools, and modify codebases with limited human oversight, making safety and trust essential goals~\cite{nvidia2025safetyrecipe,hua2024trustagent}.

One key research direction is building agents that can better understand and follow what users actually want, even when instructions are vague or incomplete. Current systems rely heavily on natural language prompts, which can be ambiguous. Future work could focus on grounding agent behavior in more structured forms of input, such as constraints, test cases, or high-level goals, that are easier to validate and reason about~\cite{fakhoury2024llm,huang2023agentcoder,li2025structured}.
Another interesting idea is to design a structured language that developers can use to express their intent clearly. This language would serve as a kind of programming interface for LLMs, allowing users to define what the agent should do, what it must avoid, and what counts as a valid solution. For example, a developer might specify that a function’s output must remain the same after refactoring, and the agent would only explore changes that meet this requirement. This kind of structure could also make it easier to apply lightweight verification tools, such as static analyzers or type checkers, to ensure the agent's suggestions are safe and correct~\cite{codeql2020,li2024iris,nunez2024autosafecoder}.

In safety-critical domains like healthcare, finance, or automotive software, agents will also need to follow strict coding standards, legal rules, and certification guidelines. Agents could be trained to recognize such constraints or be paired with rule-based systems to flag violations and suggest compliant alternatives~\cite{ouyang2022training,he2024instruction}.
Another important challenge is ethical alignment. Because agentic systems are trained on large and diverse codebases, they may learn unsafe or biased practices~\cite{liu2023uncovering,pearce2025asleep,huang2024bias}. Research efforts are needed to detect and reduce risks such as generating insecure code, leaking sensitive data, or reinforcing stereotypes. Techniques from responsible AI, like behavioral audits, adversarial testing, and human feedback, can be adapted to this setting.

To build user trust, agentic programming systems should also be able to explain their decisions~\cite{rajani2019explain,creswell2022faithful,DBLP:conf/nips/ShinnCGNY23}. Developers want to know why an agent made a change, where the idea came from, and what trade-offs are involved. Research on explanation generation, source citation, and visualization of code changes will help make LLM behavior more transparent and understandable~\cite{liu2024reliability,sun2022investigating,huang2023citation}.
Agents should also be aware of their limitations. When they are uncertain, they should be able to flag their confidence level~\cite{kuhn2023semantic,creswell2022faithful}, suggest multiple options \cite{DBLP:conf/nips/MadaanTGHGW0DPY23, DBLP:conf/nips/ShinnCGNY23}, or ask the user for confirmation \cite{mu2024clarifygpt,zhang2024modeling}. Designing agents that can adapt their behavior based on task difficulty or user feedback is a promising direction \cite{DBLP:conf/nips/ShinnCGNY23,sun2023adaplanner,damani2024learning,manvi2024adaptive,chen2024magicore}.
Furthermore, safety mechanisms like undo and audit trails will be essential. Agents should keep track of what they have changed and allow users to roll back actions easily~\cite{patil2024goex,jiang2024rocode}. Future work could explore automatic snapshotting, reversible code edits, and tight integration with version control systems to support safe collaboration.

\subsection{Multi-agent Collaboration\label{subsec:future-Multi}}

As coding agents increasingly operate in teams and interact with heterogeneous services, effective multi-agent collaboration requires standardized communication protocols that provide low-latency, scalable, and semantics-aware coordination. However, current solutions remain limited and fall short of these needs.
Future communication protocols should move beyond ad-hoc adapters and heterogeneous web services toward native support for multi-agent collaboration. Promising directions include the design of lightweight, low-latency group communication frameworks that natively support both agent--to--agent and agent--to--tool interactions, while offering standardized session management for dynamic participation and coordination~\cite{jie_2012TSC,A2A,Eurosys25_verma}. Recent standards efforts such as IEEE P3394 propose a universal framework with session, channel, and message abstractions, enabling consistent semantics and interoperability across diverse transports~\cite{tong2025ieee}. Building on these initiatives, future research should emphasize semantics-aware routing, conflict resolution, and consistency guarantees in distributed multi-agent settings. Ultimately, advancing collaboration-centric protocols will be essential for scaling efficient agentic systems in practice.

\subsection{System Support for AI Coding Agents\label{subsec:future-System}}
Realizing agentic programming at scale also requires strong systems-level support. When offered as a cloud service~\cite{aws_sagemaker,azure_ml,kserve,huggingface}, LLMs and coding agents must be backed by infrastructures that can efficiently manage resources, communication, and state.
 Agentic workflows combine code generation, compilation, testing, and tool orchestration, with highly variable resource demands~\cite{yang2024swe,wang2024openhands}. Systems must schedule CPU, GPU, and memory adaptively, balance throughput across users, and minimize latency for interactive tasks~\cite{yu2021gillis,fu2024serverlessllm,yu2022orca,patel2024splitwise,aminabadi2022deepspeed}. 
 During the iterative code generation and testing process, agents are likely to exchange large volumes of data with tools like compilers, debuggers, profilers, and repositories. Reducing overhead through caching~\cite{kwon2023efficient,liu2024cachegen}, compression~\cite{yang2024pyramidinfer,jiang2023llmlingua,hooper2024kvquant,zhao2024atom}, and lightweight protocols~\cite{grpc2015,moritz2018ray} is essential for efficiency in distributed environments.
Equally important is context and state management. To support interruption, resumption, and collaboration, systems must persist not only source code but also goals, annotations, and intermediate representations. Efficient checkpointing and synchronization are key enablers here.
Finally, multi-agent collaboration introduces distributed systems challenges: coordinating tasks, resolving conflicts, and ensuring consistency when multiple agents - or human–agent teams - work concurrently on shared codebases.

\section{Conclusion}
\label{sec:conclusion}
We have presented a comprehensive review of AI agentic programming. This new software paradigm, driven by the recent success of LLMs, is a transformative shift in how software can be created, maintained, and evolved. By combining the capabilities of LLMs with planning, tool use, and iterative refinement, coding agents are beginning to automate complex, multi-stage programming workflows that traditionally required significant human involvement. These systems can not only generate and test code but also interact with development tools, decompose tasks, and adapt based on feedback, bringing us closer to the vision of autonomous software development.

We have introduced a taxonomy of AI coding agents and architectures, reviewed underpinning techniques like context management and tool integration, and discussed how existing benchmarks evaluate the capabilities of coding agents. We have summarized the progress made in enabling LLM-based AI agents to reason over tasks, interface with software development tools, and operate in increasingly sophisticated ways. At the same time, we have identified multiple open challenges that must be addressed to ensure these systems are safe, reliable, and usable in real-world settings. These include limitations in context handling, the need for persistent and structured memory, alignment with user intent, human-AI collaboration, and verification of agent behavior. As software developers are increasingly relying on AI coding agents, these concerns will become more pressing and will require interdisciplinary solutions drawing from programming languages, human-computer interaction, software engineering, and responsible AI.

Looking ahead, AI agentic programming offers exciting opportunities to fundamentally rethink the software development practice. Whether as collaborative partners in interactive workflows or as autonomous systems that manage long-running tasks, these agents have the potential to augment developer productivity, reduce software maintenance costs, and expand access to programming. We hope this survey serves as a foundation for researchers and practitioners to navigate the emerging landscape of AI agentic programming and to accelerate progress in building the next generation of intelligent and trustworthy software development tools.

\bibliographystyle{ACM-Reference-Format}
\bibliography{references}

\end{document}